\documentclass[journal=achre4,manuscript=article,layout=onecolumn]{achemso}

\usepackage[version=3]{mhchem} 
\usepackage{amsfonts}
\usepackage{amssymb,amsmath}
\usepackage{graphicx}
\usepackage{subfigure}
\usepackage{threeparttable}
\usepackage{booktabs}
\usepackage[linesnumbered,boxed]{algorithm2e}
\usepackage{enumerate}
\usepackage{colortbl}
\usepackage{bbm} 

\title[\texttt{achemso} demonstration]
{NAC-TDDFT: Time-dependent density functional theory for nonadiabatic couplings}

\author{Zikuan Wang, Chenyu Wu and Wenjian Liu}
\email{liuwj@sdu.edu.cn}
\affiliation{Qingdao Institute for Theoretical and Computational Sciences, Institute of Frontier and Interdisciplinary Science, Shandong University, Qingdao, Shandong 266237, China}

\begin{document}

\begin{abstract}
First-order nonadiabatic coupling matrix elements (fo-NACMEs) are the basic quantities in theoretical descriptions of
electronically nonadiabatic processes that are ubiquitous in molecular physics and chemistry.
Given the large size of systems of chemical interests, time-dependent density functional theory (TDDFT) is usually the first choice. However,
the lack of wave functions in TDDFT renders the formulation of NAC-TDDFT for fo-NACMEs conceptually difficult. The present account
aims to analyze the available variants of NAC-TDDFT in a critical but concise manner and meanwhile point out the proper ways for implementation.
It can be concluded, from both theoretical and numerical points of view, that the equation of motion-based variant of NAC-TDDFT is the right choice.
Possible future developments of this variant are also highlighted.
\end{abstract}

\section{Introduction}
Electronically nonadiabatic processes involving more than one Born-Oppenheimer (BO) potential energy surfaces (PES) are ubiquitous in chemistry, biology and materials science.
There exist two mechanisms for these to happen, purely electronic and finite nuclear mass effects. The former refers to spin-orbit couplings (SOC) responsible for
transitions between electronic states of differen spins, whereas the latter to derivative couplings causing transitions between electronic states of the same spin.
To see the latter, we start with the (clamped nuclei) electronic Schr\"odinger equation
\begin{eqnarray}
H_{\mathrm{e}}|\Psi_I(\{\mathbf{r}_i\}; \{\mathbf{R}_A\})\rangle&=&E_I(\{\mathbf{R}_A\})|\Psi_I(\{\mathbf{r}_i\}; \{\mathbf{R}_A\})\rangle,\\
H_{\mathrm{e}}&=&T_{\mathrm{e}}  + V_{\mathrm{nn}}(\{\mathbf{R}_A\}) + V_{\mathrm{ne}}(\{\mathbf{r}_i\}; \{\mathbf{R}_A\}) + V_{\mathrm{ee}}(\{\mathbf{r}_i\}), \label{He}
\end{eqnarray}
where the semi-colons emphasize that the nuclear coordinates $\{\mathbf{R}_A\}$ are parameters instead of variables like the electronic coordinates $\{\mathbf{r}_i\}$.
The eigenvalues $E_I(\{\mathbf{R}_A\})$ form PESs on which the nuclei move. Since $\{|\Psi_I(\{\mathbf{r}_i\})\rangle\}$ form a complete basis set (CBS), the total wave function
can be expanded as
\begin{eqnarray}
\Phi(\{\mathbf{r}_i\},\{\mathbf{R}_A\})&=& \sum_J \Theta_J(\{\mathbf{R}_A\}) \Psi_J(\{\mathbf{r}_i\}; \{\mathbf{R}_A\}),\label{BO_expansion}
\end{eqnarray}
which can be inserted into the full Schr\"odinger equation
\begin{eqnarray}
(T_{\mathrm{n}}+H_{\mathrm{e}})\Phi(\{\mathbf{r}_i\},\{\mathbf{R}_A\})&=&E_{\mathrm{tot}}\Phi(\{\mathbf{r}_i\},\{\mathbf{R}_A\})
\end{eqnarray}
to obtain the nuclear Schr\"odinger equation
\begin{eqnarray}
\left(T_{\mathrm{n}} + E_{I}(\{\mathbf{R}_A\})\right) \Theta_I(\{\mathbf{R}_A\}) + \sum_J H^{\mathrm{BO}}_{IJ}\Theta_J(\{\mathbf{R}_A\}) = E_{\mathrm{tot}}\Theta_I(\{\mathbf{R}_A\}), \label{NACorigin}
\end{eqnarray}
where
\begin{eqnarray}
H^{\mathrm{BO}}_{IJ} &=& - \sum_{\xi} \frac{1}{M_{\xi}} (g_{IJ}^{\xi}\partial_{\xi} + h_{IJ}^{\xi}),\quad \partial_{\xi} = \frac{d}{d \xi}\\
g_{IJ}^{\xi} &=& \langle \Psi_I | \partial_{\xi}| \Psi_J \rangle, \label{foNACME}\\
h_{IJ}^{\xi} &=& \frac{1}{2} \langle \Psi_I | \partial_{\xi}^2| \Psi_J \rangle. \label{soNACME}
\end{eqnarray}
Here, $\xi$ runs over all nuclear degrees of freedom and $M_\xi$ is the corresponding atomic mass.
Clearly, the matrix operator $H^{\mathrm{BO}}_{IJ}$ vanishes in the limit of infinite nuclear masses.
Conversely, for the true atomic masses, the operator will induce transitions between electronic states of the same spin,
especially when the states are energetically adjacent (cf. Eq. \eqref{Exact2} below).
Since the second-order nonadiabatic coupling matrix elements (NACME) \eqref{soNACME}
can be ``transformed away'' (without approximation for the considered manifold
of PESs) by replacing the canonical with the kinematic nuclear momentum\cite{AdiabaticMM}, only the first-order (fo) NACMEs \eqref{foNACME} are relevant.
In view of the relation
\begin{eqnarray}
\langle\Psi_I|[ \partial_{\xi}, H_{\mathrm{e}}]|\Psi_J\rangle=\omega_{JI}g_{IJ}^{\xi}, \quad \omega_{JI}=E_J-E_I,\quad  \forall I\ne J,\label{Exact}
\end{eqnarray}
the fo-NACMEs can be calculated as
\begin{eqnarray}
g_{IJ}^{\xi}&=&\frac{\langle\Psi_I|V_{\mathrm{ne}}^{\xi}|\Psi_J\rangle}{\omega_{JI}}=-g_{JI}^{\xi},\quad V_{\mathrm{ne}}^{\xi}=[\partial_{\xi}, V_{\mathrm{ne}}]=(\partial_{\xi} V_{\mathrm{ne}}) \label{Exact2}\\
&=&\frac{1} {\omega_{JI}} \sum_{pq} \langle \psi_p|V_{\mathrm{ne}}^{\xi}|\psi_q\rangle \gamma^{IJ}_{pq}, \quad \gamma^{IJ}_{pq}=\langle \Psi_I|a_p^\dag a_q|\Psi_J\rangle. \label{HFey_rho}
\end{eqnarray}
Use of the above Hellmann-Feynman-like expression was first made by Chernyak and Mukamel\cite{ChernyakMukamelNAC} for the fo-NACMEs between the ground and excited states (\emph{ge})
in the context of time-dependent density functional theory (TDDFT), which does provide reduced transition density matrices (rTDM) $\boldsymbol{\gamma}^{IJ}$ between a pair of states.
However, even ignoring the fact that the expression \eqref{HFey_rho} is rooted in (exact) wave function theory (WFT) and hence incompatible with TDDFT,
it is hardly useful in practice, for it requires a CBS.
For atom-centered basis sets, the convergence is extremely slow, which can easily be understood
by noticing that, as a real-space function,
$V_{\mathrm{ne}}^{\xi}$  has a dipole symmetry and behaves as $R^{-2}$ in the vicinity of the nucleus. As such,
not only steep $p$ functions but also functions of angular momentum $l+1$ have to be added to
standard basis functions of angular momentum $l$\cite{SendFurcheNAC}.
A more useful and simple approach is the finite difference approximation via fast evaluation of the overlap integrals between electronic wave functions
at displaced geometries\cite{OI-td2019,OI-NACME2021}. However, in total $6N_{\mathrm{atom}}$ evaluations of the overlap integrals are needed
to get the fo-NACMEs and care has to be taken of the dependence on geometric displacements as well as the phase alignment of the wave functions.
Therefore, analytic NAC-TDDFT should be formulated (see Sec. \ref{sec:TDDFT_NACME}) and recast into practically useful forms
(see Sec. \ref{sec:Lagrangian}). Some
numerical results will be provided in Sec. \ref{sec:application} before concluding the Account in Sec. \ref{sec:conclusion}.

The following convention is to be used: $\{i,j,k,l\}$, $\{a,b,c,d\}$ and $\{p,q,r,s\}$ denote occupied, virtual and arbitrary molecular orbitals (MO),
whereas Greek letters refer to atomic orbitals (AO).
\section{NAC-TDDFT} \label{sec:TDDFT_NACME}
Under the adiabatic approximation, TDDFT amounts to solving the following eigenvalue problem\cite{Casida1995}
\begin{equation}
\mathbf{E}\mathbf{t}_I = \omega_I\mathbf{S}\mathbf{t}_I,\quad \mathbf{t}_I^\dag\mathbf{S}\mathbf{t}_J = \delta_{IJ}, \label{Casida}
\end{equation}
where
\begin{equation}
\mathbf{E} = \left(\begin{array}{cc} \mathbf{A} & \mathbf{B} \\ \mathbf{B} & \mathbf{A}\end{array}\right), \quad
\mathbf{S} = \left(\begin{array}{cc} \mathbf{I} & \mathbf{0} \\ \mathbf{0} & -\mathbf{I} \end{array}\right), \quad
\mathbf{t}_I = \left(\begin{array}{c} \mathbf{X}_I \\ \mathbf{Y}_I \end{array}\right), \label{Emat}
\end{equation}
\begin{eqnarray}
A_{ia\sigma,jb\tau} & = & \delta_{\sigma\tau}(\delta_{ij}F_{ab\sigma} - \delta_{ab}F_{ji\sigma}) + K_{ia\sigma{}jb\tau}, \label{UTD-Amat}\\
B_{ia\sigma,jb\tau} & = & K_{ia\sigma{}bj\tau}, \\
K_{pq\sigma,rs\tau} & = & (pq\sigma|sr\tau)  - c_{\mathrm{x}}\delta_{\sigma\tau}(pr\sigma|sq\sigma)+ c_{\mathrm{xc}}f_{pq\sigma{}sr\tau}^{\mathrm{xc}}[\rho],\label{Kmat}
\end{eqnarray}
\begin{equation}
F_{\mu\nu\sigma} = h_{\mu\nu} + \sum_{i\tau}[(\mu\sigma \nu\sigma|i\tau i\tau)-c_{\mathrm{x}}\delta_{\sigma\tau}(\mu\sigma i\tau|i\tau \nu\sigma)] + c_{\mathrm{xc}}v_{\mu\nu\sigma}^{\mathrm{xc}}[\rho], \quad h_{\mu\nu} = T_{\mu\nu} + (V_{\mathrm{ne}})_{\mu\nu}. \label{Fao}
\end{equation}
Here, $\sigma$ and $\tau$ are spin indices.
One major issue here is that the eigenvectors of Eq. \eqref{Casida} do not correspond to excited-state wave functions, but
 are related to the TDM describing linear response of the ground state density to an external field
(cf. equation (103) in Ref. \citenum{RTDDFTrev}). The lack of wave functions
renders the formulation of TDDFT for fo-NACMEs at least formally difficult.
In the next subsections, the auxiliary/pseudo wave function (AWF), equation of motion (EOM), and time-dependent response/perturbation theory (TDPT) based formulations are
analyzed sequentially.


\subsection{AWF-based formulation} \label{sec:auxiliary}
To draw analogy with WFT, we start with the Tamm-Dancoff approximation (TDA) to Eq. \eqref{Casida},
\begin{equation}
\mathbf{A}\mathbf{X}_I = \omega_I\mathbf{X}_I.\label{TDAeq}
\end{equation}
If Hartree-Fock (HF) is used as the functional,
the $\mathbf{A}$ matrix \eqref{UTD-Amat} is just the Hamiltonian matrix in the manifold of singly excited configurations $\{|\Psi_i^a\rangle = a_a^\dag a_i|\Psi_{\mathrm{HF}}\rangle\}$.
Therefore, in this case, the eigenvector $\mathbf{X}_I$ represents simply the
coefficients of the CIS (configuration interaction singles) wave function $\Psi^{\mathrm{CIS}}_I$,
\begin{eqnarray}
|\Psi^{\mathrm{CIS}}_I\rangle=\sum_{ia}|\Psi_i^a\rangle (\mathbf{X}_I)_{ia}.
\end{eqnarray}
Extending the above to an arbitrary functional leads to
\begin{equation}
|\Psi^{\mathrm{TDA}}_I\rangle = \sum_{ia} (\mathbf{X}_I)_{ia} a_a^\dag a_i|\Psi_{\mathrm{KS}}\rangle,
\end{equation}
which can be termed ``auxiliary wave function'' (AWF).
In this way, the \emph{ge} and \emph{ee} (excited state-excited state) fo-NACMEs can readily be obtained as
\begin{eqnarray}
g_{0I}^{\xi} &=& \sum_{ia} \langle \psi_i | \partial_{\xi} | \psi_a \rangle (\mathbf{X}_I)_{ia}, 
\label{CIS_0I}\\
g_{IJ}^{\xi} &=& \sum_{iab} (\mathbf{X}_I)_{ia} (\mathbf{X}_J)_{ib} \langle \psi_a | \partial_{\xi} | \psi_b \rangle - \sum_{ija} (\mathbf{X}_I)_{ia} (\mathbf{X}_J)_{ja} \langle \psi_j | \partial_{\xi} | \psi_i\rangle.\label{CIS_IJ}
\end{eqnarray}
However, these are not yet the final working equations, as they involve nuclear derivatives of the MOs,
which must be either calculated directly from coupled-perturbed KS (CPKS) equations or eliminated by the Z-vector approach (see Sec. \ref{sec:Lagrangian}).

The situation becomes very different for the full TDDFT, which has no direct analogy with any WFT.
One common practice is to construct a CIS-like wave function that reproduces some property (e.g.,
electric polarizability) in a sum-of-states form, and then feed it to Eq. (\ref{foNACME}). Different choices of properties
then lead to different expressions, e.g.,
\begin{equation}
g_{0I}^{\xi} = \sum_{ia} (\mathbf{X}_I + \mathbf{Y}_I)_{ia} \langle \psi_i | \partial_{\xi} | \psi_a \rangle,\label{Tavern-fo-NACME}
\end{equation}
and
\begin{equation}
g_{0I}^{\xi} = \sum_{ia} (\mathbf{X}_I -\mathbf{Y}_I)_{ia} \langle \psi_i | \partial_{\xi} | \psi_a \rangle,\label{Hu-fo-NACME}
\end{equation}
were proposed by Tavernelli\cite{Tavernelli2007} and Hu\cite{Hu2007NAC, Hu2010NAC}, respectively,
for the \emph{ge} fo-NACMEs. Even more choices are possible for the \emph{ee} fo-NACMEs\cite{Tavernelli2010, Subotnik2015, Tavernelli2007}.
Due to the lack of theoretical rigor, there is no \emph{a priori} argument to favor one over another. In particular,
there is no \emph{a priori} guarantee that any of them is exact 
if both the functional and kernel were exact. It is shown in the next subsections that NAC-TDDFT
can indeed be formulated  more properly\cite{lzdNACtheory,lzdNACimplementation,PAW_NAC,FurcheSubotnik2015,Herbert2015NAC}.

\subsection{EOM-based formulation}
In the EOM formalism,\cite{EOM} one defines an excitation operator $O_I^\dag$ that promotes the ground state to an excited state
\begin{equation}
| I \rangle = O_I^\dag | 0 \rangle,
\end{equation}
and is subject to the killer condition
\begin{equation}
\langle 0|O_I^\dag = 0. \label{killer}
\end{equation}
After determining $O_I^\dag$ by the EOM
\begin{equation}
 \frac{1}{2} \langle 0 | [\delta O_I, [H_{\mathrm{e}}, O_I^\dag ] ]+[[ \delta O_I, H_{\mathrm{e}}], O_I^\dag] | 0 \rangle = \omega_I \langle 0 | [\delta O_I, O_I^\dag] | 0 \rangle,\quad \omega_I=E_I-E_0, \label{Casida_EOM}
\end{equation}
the \emph{ge} and \emph{ee} matrix elements of an arbitrary operator $O$ can be calculated as
\begin{eqnarray}
\langle 0 | O | I \rangle &=& \langle 0 | O O_I^\dag | 0 \rangle \label{operator_EOM_0I}\\
                          &=& \langle 0 | [ O, O_I^\dag ] | 0 \rangle, \label{operator_EOM_0I_commutator}\\
\langle I |O | J \rangle &=& \langle 0 | O_I O O_J^\dag | 0 \rangle \label{operator_EOM_IJ}\\
                         &=& \frac{1}{2} \langle 0 | [ O_I, [O, O_J^\dag] ] - [ [O, O_I], O_J^\dag ] | 0 \rangle, \quad I \neq J. \label{operator_EOM_IJ_commutator}
\end{eqnarray}
Note that use of the killer condition \eqref{killer} has been made when going from Eq. \eqref{operator_EOM_0I}/\eqref{operator_EOM_IJ} to \eqref{operator_EOM_0I_commutator}/\eqref{operator_EOM_IJ_commutator},
to be consistent with the structure of Eq. \eqref{Casida_EOM}.

Consider, e.g., the random phase approximation (RPA) with
\begin{equation}
O_I^\dag = \sum_{ia} (\mathbf{X}_I)_{ia} a_a^\dag a_i - (\mathbf{Y}_I)_{ia} a_i^\dag a_a. \label{OI}
\end{equation}
It can readily be checked that Eq. \eqref{Casida_EOM} with \eqref{OI} is just Eq. \eqref{Casida} with $c_{\mathrm{x}} = 1$ and $c_{\mathrm{xc}} = 0$,
which can simply be translated to TDDFT with a functional other than HF. Note, however, the $O_I^\dag$ operator \eqref{OI} does
not satisfy the killer condition due to the presence of $a_i^\dag a_a$. As such, in principle Eqs. \eqref{operator_EOM_0I} and \eqref{operator_EOM_IJ}
should be used for the RPA/TDDFT \emph{ge} and \emph{ee} matrix elements, respectively. The resulting expressions are then the same as those by the previous CIS/TDA-based approach.
The very trick for nonvanishing deexcitation amplitudes $\mathbf{Y}_I$
is still to use Eqs. \eqref{operator_EOM_0I_commutator} and \eqref{operator_EOM_IJ_commutator}.
Straightforward derivations then yield
\begin{eqnarray}
\langle I | O | J \rangle &=& \langle I | O | J \rangle_0 + \langle I | O | J \rangle_1,\label{EOM_IJ}\\
\langle I | O | J \rangle_0&=&\sum_{pq} \tilde{\gamma}^{IJ}_{pq} \langle \psi_p | O | \psi_q \rangle, \label{EOM_IJ0}\\
\langle I | O | J \rangle_1&=& \frac{1}{2} \langle 0 | [ O_I, [O, O_J^\dag]_1 ] - [ [O, O_I]_1, O_J^\dag ] | 0 \rangle, \label{EOM_IJ1}\\
\protect[O, O_J^\dag\protect]_1 &=& \sum_{ia} [O, (\mathbf{X}_I)_{ia}] a_a^\dag a_i - [O, (\mathbf{Y}_I)_{ia}] a_i^\dag a_a. 
\end{eqnarray}
where $\langle I | O | J \rangle_1$ appears only if $O$ contains nuclear derivatives, and the
 \emph{ge} and \emph{ee} TDMs  are defined in Eqs. \eqref{ge-gammaEOM} and \eqref{ee-gammaEOM}, respectively,
\begin{equation}\label{ge-gammaEOM}
\tilde{\gamma}^{0I}_{ia} = (\mathbf{X}_I)_{ia}, \quad
\tilde{\gamma}^{0I}_{ai} = (\mathbf{Y}_I)_{ia}, \quad
\tilde{\gamma}^{0I}_{ij} = \tilde{\gamma}^{0I}_{ab} = 0,
\end{equation}
\begin{subequations}\label{ee-gammaEOM}
\begin{equation}
\tilde{\gamma}^{IJ}_{ij} = - \sum_{a}\left[(\mathbf{X}_J)_{ia} (\mathbf{X}_I)_{ja} + (\mathbf{Y}_I)_{ia} (\mathbf{Y}_J)_{ja}\right],
\end{equation}
\begin{equation}
\tilde{\gamma}^{IJ}_{ab} = \sum_{i}\left[(\mathbf{X}_I)_{ia} (\mathbf{X}_J)_{ib} + (\mathbf{Y}_J)_{ia} (\mathbf{Y}_I)_{ib}\right],
\end{equation}
\begin{equation}
\tilde{\gamma}^{IJ}_{ia} = \tilde{\gamma}^{IJ}_{ai} = 0.
\end{equation}
\end{subequations}

Some remarks are in order.
\begin{enumerate}[(1)]
\item It is the use of the nested commutators of the form $[ O_I, [O, O_J^\dag] ]$ that allows the deexcitation amplitudes $\mathbf{Y}_I$ to contribute to
the matrix elements of $O$. Such commutators are only present in EOM but not in the Schr\"odinger equation itself,
which explains the difficulty of formulating a unique AWF for TDDFT, except for the TDA variant.


\item The \emph{ge} transition matrix element $\langle 0 | O | I \rangle_0$ defined by Eqs. \eqref{EOM_IJ0} and \eqref{ge-gammaEOM} can be simplified to
\begin{equation}
\langle 0 | O | I \rangle_0 = \sum_{ia} (\mathbf{X}_I + \mathbf{Y}_I)_{ia} \langle \psi_i | O | \psi_a \rangle
\end{equation}
if $O$ is real Hermitian or to
\begin{equation}
\langle 0 |O | I \rangle_0 = \sum_{ia} (\mathbf{X}_I - \mathbf{Y}_I)_{ia} \langle \psi_i | O | \psi_a \rangle\label{geEOMfoNACME}
\end{equation}
if $O$ is real anti-Hermitian.
The important corollary is that, in the AWF-based approach, one should construct the AWF with $\mathbf{X}_I + \mathbf{Y}_I$
in the case of $O = V_{\mathrm{ne}}^{\xi}$ or $ \mathbf{r}$ but with $\mathbf{X}_I - \mathbf{Y}_I$ in the case of $O = \partial_{\xi}$.
This is related to the well-known fact that $\mathbf{A} + \mathbf{B}$ serves as the orbital Hessian for responses to a real (``electric'') perturbation,
whereas $\mathbf{A} - \mathbf{B}$ for responses to a purely imaginary (``magnetic'') perturbation (which is related to real anti-Hermiticity scaled by the imaginary factor $\mathbbm{i}$).
As such, TDDFT does not admit a unique AWF\cite{etienne2021auxiliary,Luzanov2010}, as the form of the AWF depends on the
type of perturbation studied. The AWF constructed to reproduce the dipole polarizability (i.e., $\mathbf{X}_I + \mathbf{Y}_I$) gives rise to incorrect \emph{ge} fo-NACMEs in
view of Eq. \eqref{geEOMfoNACME} with $O=\partial_{\xi}$, whereas the same AWF used for $\langle \Psi_I | V_{\mathrm{ne}}^{\xi} | \Psi_J \rangle$ in
Eq. \eqref{HFey_rho} will yield the correct fo-NACMEs, but only in the CBS limit.

\item None of the AWF-based methods can reproduce the \emph{ee} fo-NACMEs obtained by EOM, unless further approximations are made\cite{Subotnik2015}. For example, starting with an AWF $\mathbf{X}_I \pm \mathbf{Y}_I$, one obtains
\begin{equation}
\langle I | O | J \rangle_0 = \sum_{iab} (\mathbf{X}_I \pm \mathbf{Y}_I)_{ia} (\mathbf{X}_J \pm \mathbf{Y}_J)_{ib} \langle \psi_a | O | \psi_b \rangle
- \sum_{ija} (\mathbf{X}_I \pm \mathbf{Y}_I)_{ia} (\mathbf{X}_J \pm \mathbf{Y}_J)_{ja} \langle \psi_j | O | \psi_i \rangle,
\end{equation}
which involves cross terms like $(\mathbf{X}_I)_{ia} (\mathbf{Y}_J)_{ib}$ not present in the EOM expression (see Eqs. \eqref{EOM_IJ0} and \eqref{ee-gammaEOM}).

\item The second term in Eq. \eqref{EOM_IJ} with $O=\partial_{\xi}$ has been derived before\cite{lzdNACtheory} and is hence not repeated here.
The final expressions for the \emph{ge} and \emph{ee} fo-NACMEs read
\begin{eqnarray}
\langle 0 |\partial_{\xi} | I \rangle &=& \sum_{ia} (\mathbf{X}_I - \mathbf{Y}_I)_{ia} \langle \psi_i |\partial_{\xi}| \psi_a \rangle, \label{NAC_EOM_0I}\\
\langle I |\partial_{\xi} | J \rangle &=&  \sum_{pq}\tilde{\gamma}^{IJ}_{pq} \langle \psi_p |\partial_{\xi} | \psi_q \rangle + \omega_{JI}^{-1} \mathbf{t}_I^\dag
(\partial_{\xi}\mathbf{E})\mathbf{t}_J. \label{NAC_EOM_IJ}
\end{eqnarray}
Clearly, the AWF-based expression \eqref{Hu-fo-NACME}\cite{Hu2007NAC, Hu2010NAC} happens to be correct but the expression
\eqref{Tavern-fo-NACME}\cite{Tavernelli2007} should be rejected.

\end{enumerate}

Albeit elegant, the above EOM-based formulation of fo-NACMEs is still not satisfactory, for the killer condition \eqref{killer} does not hold strictly.
By contrast, the TDPT-based formulation\cite{lzdNACtheory} of NAC-TDDFT is more rigorous, as detailed in the next subsection.

\subsection{TDPT-based formulation}
Apart from the previous time-independent AWF and EOM formulations, NAC-TDDFT can also be formulated\cite{lzdNACtheory,lzdNACimplementation} through TDPT,
which describes the changes of the ground state to a time-dependent perturbation
\begin{equation}
V(t) = \sum_k e^{-\mathbbm{i}\omega_k t} \sum_\beta V^\beta (\omega_k) \epsilon_\beta (\omega_k),
\end{equation}
where $V^\beta (\omega_k)$ is the perturbation operator with strength $\epsilon_\beta (\omega_k)$. Under this perturbation,
the ground state is also dependent on time, i.e., $| 0 \rangle = | 0(t) \rangle$. Since
we are only interested in the evolution of a ground state property, e.g., $\langle 0(t) | V^\alpha | 0(t) \rangle$ with $V^\alpha$ being
time-independent,  we can separate out the global phase from $| 0(t) \rangle$,
\begin{equation}
| 0(t) \rangle = e^{\mathbbm{i}\theta(t)} | \tilde{0}(t) \rangle, \quad \theta(t)|_{V(t) \equiv 0} = -tE_0(\{\mathbf{R}_A\}).
\end{equation}
We then define
\begin{equation}
A(t) = \langle \tilde{0}(t) | V^\alpha | \tilde{0}(t) \rangle,\label{Adef}
\end{equation}
which can be expanded as
\begin{eqnarray}
A(t) &=& A^{(0)}(t) + A^{(1)}(t) + A^{(2)}(t) + \cdots,\label{Aexpan}\\
A^{(0)}(t) &=& \langle 0 | V^\alpha | 0 \rangle,\\
A^{(1)}(t) &=& \sum_k e^{-\mathbbm{i}\omega_k t} \sum_\beta A^\beta (\omega_k) \epsilon_\beta(\omega_k),\\
A^{(2)}(t) &=& \sum_{kl} e^{-\mathbbm{i}(\omega_k + \omega_l) t} \sum_{\beta\gamma} A^{\beta\gamma} (\omega_k, \omega_l) \epsilon_\beta(\omega_k) \epsilon_\gamma(\omega_l).
\end{eqnarray}
To derive the linear [$A^\beta (\omega_k)$] and quadratic [$A^{\beta\gamma} (\omega_k, \omega_l)$] response functions in the framework of TDDFT, we parameterize the phase-separated state
$| \tilde{0}(t) \rangle$ as
\begin{equation}
| \tilde{0}(t) \rangle =e^{-\kappa(t)} | 0 \rangle, \quad \kappa(t) = -\kappa^\dag(t), \label{UI}
\end{equation}
where $\kappa(t)$  can be expanded as
\begin{equation}
\kappa(t) = \kappa^{(1)}(t) + \kappa^{(2)}(t) + \cdots,\quad \kappa^{(0)}(t)=0
\end{equation}
in accordance with Eq. \eqref{Aexpan}. To determine $\kappa(t)$, we invoke
Ehrenfest's theorem\cite{Ehrenfest_theorem}
\begin{equation}
\langle 0 | [ \hat{q}, e^{\kappa(t)} (H_{\mathrm{KS}}(t) + V(t) - \mathbbm{i} \frac{\partial}{\partial t}) e^{-\kappa(t)} ] | 0 \rangle = 0, \label{Ehrenfest}
\end{equation}
where $H_{\mathrm{KS}}(t)$ is the KS operator and $\hat{q}$ is an arbitrary operator in the operator basis used to expand $O_I^\dag$,
taken here to be $a_p^\dag a_q$. Both Eq. \eqref{Adef} in conjunction with Eq. \eqref{UI} and Eq. \eqref{Ehrenfest} possess operators of the form $e^{\kappa(t)} O e^{-\kappa(t)}$, which can be expanded as
\begin{equation}
e^{\kappa(t)} O e^{-\kappa(t)} = O + [ \kappa(t), O ] + \frac{1}{2}[ \kappa(t), [ \kappa(t), O ] ] + \cdots.\label{BCH}
\end{equation}
It is interesting to note that, if only the first-order term of $\kappa(t)$ is retained,
applying the expansion \eqref{BCH} to Eq. (\ref{Ehrenfest}) will yield the EOM \eqref{Casida_EOM}, which is equivalent to
the standard TDDFT \eqref{Casida}. Likewise, repeating the same to Eq. \eqref{Adef} will recover the commutator type of metrics \eqref{operator_EOM_0I_commutator} and \eqref{operator_EOM_IJ_commutator},
thereby explaining the underlying reasoning for EOM to take such metrics.
However, $\kappa^{(2)}(t)$ also contributes to the \emph{ee} fo-NACMEs, which is missed by the linear-response EOM \eqref{Casida_EOM}.
The contribution of $\kappa^{(2)}(t)$ is to be determined by the quadratic response equation
\begin{equation}
(\mathbf{E} - \omega_{JI} \mathbf{S}) \mathbf{t}_{IJ} = \mathbf{V}_{IJ}, \quad
\mathbf{t}_{IJ} = \left(\begin{array}{c} \mathbf{X}_{IJ} \\ \mathbf{Y}_{IJ} \end{array}\right), \label{quadresp}
\end{equation}
with $\mathbf{V}_{IJ}$ defined by equation (32) of Ref. \citenum{lzdNACimplementation}. The EOM \emph{ee} fo-NACMEs \eqref{NAC_EOM_IJ} should hence be extended accordingly to
\begin{eqnarray}
\langle I |\partial_{\xi}| J \rangle &=& \sum_{pq} \bar{\gamma}^{IJ}_{pq} \langle \psi_p | \partial_{\xi} | \psi_q \rangle + \omega_{JI}^{-1} \mathbf{t}_I^\dag (\partial_{\xi}\mathbf{E})\mathbf{t}_J,\label{RTee}\\
\bar{\gamma}^{IJ}_{pq}&=&\tilde{\gamma}^{IJ}_{pq}+
(\mathbf{X}_{IJ})_{pq}\delta_{pi}\delta_{qa} - (\mathbf{Y}_{IJ})_{qp}\delta_{qi}\delta_{pa}. 
\label{TDPTrTDM}
\end{eqnarray}
It turns out that the last, quadratic-response term does improve the accuracy of the \emph{ee} fo-NACMEs but often
introduces numerical instability and even divergence\cite{lzdNACimplementation,Herbert2015NAC, FurcheSubotnik2015},
which arises whenever three TDDFT excitation energies $\omega_I, \omega_J, \omega_K$ satisfy the three-state resonance condition
$\omega_I - \omega_J = \omega_K$, thereby rendering the left-hand side of  Eq. (\ref{quadresp}) singular.
It has therefore been recommended\cite{lzdNACimplementation} that the original EOM expression \eqref{NAC_EOM_IJ} for the \emph{ee} fo-NACMEs
should be used instead. Still, however, such TDPT-based formulation puts NAC-TDDFT on a firm basis, at least conceptually.
Note in passing that this type of divergence is not limited to adiabatic TDDFT but is present in all approximate theories\cite{Furche_quad_resp_diverge}.

\section{Proper implementation} \label{sec:Lagrangian}
Having discussed the essentials of NAC-TDDFT, we comment briefly on the actual implementation. The major issue here is how to handle
the implicit dependence of the MOs on the nuclear coordinates imposed by the Brillouin and orthonormality conditions
\begin{equation}
F_{ia} = 0; \quad S_{pq} = \delta_{pq}. \label{Fia}
\end{equation}
To transform the nuclear derivatives of the MOs into a tractable form, we first note that the fo-NACMEs can be interpreted as the nuclear derivatives of some function.
For example, Eqs. \eqref{NAC_EOM_0I} and \eqref{NAC_EOM_IJ} are, respectively, the total derivatives of
\begin{equation}
g_{0I}(\{\mathbf{R}_A\}, \mathbf{C}(\{\mathbf{R}_A\})) = \sum_{ia} (\mathbf{X}_I - \mathbf{Y}_I)_{ia} \langle \psi_i(\{\mathbf{R}_A\}_0) | \psi_a(\{\mathbf{R}_A\}) \rangle, \label{g0I}
\end{equation}
and
\begin{equation}
g_{IJ}(\{\mathbf{R}_A\}, \mathbf{C}(\{\mathbf{R}_A\})) =  \sum_{pq} \tilde{\gamma}^{IJ}_{pq} \langle \psi_p(\{\mathbf{R}_A\}_0) | \psi_q (\{\mathbf{R}_A\}) \rangle +\omega_{JI}^{-1} \mathbf{t}_I^\dag \mathbf{E}(\{\mathbf{R}_A\}) \mathbf{t}_J, \label{gIJ}
\end{equation}
with respect to $\{\mathbf{R}_A\}$ at the expansion point $\{\mathbf{R}_A\}_0$.
That is, only the ket of $\langle \psi_p | \psi_q \rangle$ is to be differentiated.
The next step is to introduce a suitable Lagrangian incorporating explicitly the constraints \eqref{Fia}
\begin{eqnarray}
L[\{\mathbf{R}_A\}, \mathbf{C}(\{\mathbf{R}_A\}), \mathbf{Z}, \mathbf{W}] & = & g(\{\mathbf{R}_A\}, \mathbf{C}(\{\mathbf{R}_A\})) + \sum_{ia}Z_{ia}F_{ia}(\mathbf{C}(\{\mathbf{R}_A\})) \nonumber\\
& - & \sum_{pq} W_{pq} (S_{pq}(\mathbf{C}(\{\mathbf{R}_A\})) - \delta_{pq}), \label{Lagrangian}
\end{eqnarray}
so as to eliminate the need to differentiate the MO coefficients $\mathbf{C}(\{\mathbf{R}_A\})$ with respect to $\{\mathbf{R}_A\}$. This is achieved by
requiring the Lagrangian to be stationary with respect to $\mathbf{C}(\{\mathbf{R}_A\})$, $ \mathbf{Z}$, and $\mathbf{W}$ (but not with respect to $\mathbf{C}(\{\mathbf{R}_A\}_0)$!),
such that $\mathbf{C}(\{\mathbf{R}_A\})$ can be treated as an independent, unconstrained variational parameter rather than a parameter that depends on $\{\mathbf{R}_A\}$.
Consequently, the fo-NACMEs take simply the following form
\begin{equation}
g^\xi = \frac{d L}{d \xi} = \frac{\partial L}{\partial \xi} = \frac{\partial g}{\partial \xi} + \sum_{ia}Z_{ia}\frac{\partial F_{ia}}{\partial \xi} - \sum_{pq} W_{pq} \frac{\partial S_{pq}}{\partial \xi}, \label{dL}
\end{equation}
where the first equality arises because the last two terms of Eq. (\ref{Lagrangian}) are zero when the stationary conditions with respect to $\mathbf{Z}$ and $\mathbf{W}$ are fulfilled,
whereas the second equality is due to that the dependence of $L$ on $\{\mathbf{R}_A\}$ is completely explicit (for a detailed discussion of the Lagrangian technique, see Appendix B of Ref. \citenum{XTDDFTgrad}).

The Lagrange multipliers $\mathbf{Z}$ and $\mathbf{W}$ are determined by the stationary condition of $L$ with respect to $\mathbf{C}(\{\mathbf{R}_A\})$ (see
equations (74)-(80) in Ref. \citenum{lzdNACimplementation}). In particular, the Z-vector equation
\begin{equation}
(\mathbf{A} + \mathbf{B})\mathbf{Z} = \mathbf{X}_I - \mathbf{Y}_I \label{Zvector_0I}
\end{equation}
for the TDDFT \emph{ge} fo-NACMEs can be solved directly
\begin{equation}
\mathbf{Z} = (\mathbf{A} + \mathbf{B})^{-1}(\mathbf{X}_I - \mathbf{Y}_I) = \omega_{I}^{-1} (\mathbf{X}_I + \mathbf{Y}_I). \label{Z0I}
\end{equation}
Ironically, the TDA version of Eq. \eqref{Zvector_0I}
\begin{equation}
(\mathbf{A} + \mathbf{B})\mathbf{Z} = \mathbf{X}_I \label{Zvector_0I_TDA}
\end{equation}
has to be solved explicitly. As a result, the TDA \emph{ge} fo-NACMEs are more expensive to compute than those of full TDDFT.
Actually, the computational cost of the TDDFT \emph{ge} fo-NACMEs is very similar to that of
the DFT gradients (which also do not involve solving a Z-vector equation due to their variational nature), with only one additional complication:
the integrals $\langle \mu | \partial_{\xi} | \nu\rangle$ in $\langle \psi_p | \partial_{\xi} | \psi_q \rangle$
must be evaluated in a way that ensures translational invariance\cite{ETF}.

At this point, it deserves to be mentioned that, in the CBS limit,
the derivatives $\frac{\partial F_{ia}}{\partial \xi}$ in Eq. \eqref{dL}
reduce to $\langle \psi_i | V_{\mathrm{ne}}^{\xi} | \psi_a \rangle$,
while all other terms therein vanish, thereby reproducing the Hellmann-Feynman-like expression \eqref{HFey_rho}.
In view of Eq. (\ref{Fao}), the $V_{\mathrm{ne}}$ contribution to $\frac{\partial F_{ia}}{\partial \xi}$ reads
\begin{equation}
\frac{\partial}{\partial \xi} \langle \psi_i | V_{\mathrm{ne}} | \psi_a \rangle = \sum_{\mu\nu} C_{\mu i} \frac{\partial}{\partial \xi} \langle \mu | V_{\mathrm{ne}} | \nu \rangle C_{\nu a}.
\end{equation}
Since the basis set representation of $V_{\mathrm{ne}}$ is much easier than that of $V_{\mathrm{ne}}^\xi$, it is clear that Eq. \eqref{dL} converges much faster than Eq. \eqref{HFey_rho}. In particular, no special basis functions\cite{SendFurcheNAC} are needed, which is a point that should have been expected from the very beginning, for the fo-NACMEs are valence properties anyway.

Finally, the computational cost of the \emph{ee} fo-NACMEs is very similar to that of the TDDFT gradients\cite{FurcheTDDFTgrad,FurcheTDDFTgradErratum,XTDDFTgrad}, as already noticed before\cite{lzdNACimplementation}.
There is a simple reason for this: the second term of Eq. \eqref{gIJ} resembles closely the TDDFT excitation energy $\omega_I=\mathbf{t}_I^\dag \mathbf{E}(\{\mathbf{R}_A\}) \mathbf{t}_I$, such that
the TDDFT \emph{ee} fo-NACMEs can readily be implemented in a code that already supports TDDFT gradients.

\section{Illustration} \label{sec:application}
Azulene is known as a textbook example of anti-Kasha fluorescence. That is, the $S_2 \to S_1$
internal conversion (IC) therein is not fast enough to quench completely the $S_2 \to S_0$ fluorescence, while the $S_1 \to S_0$ IC
is sufficiently fast to make the $S_1$ state nearly non-fluorescent, thereby leading to a fluorescence dominated by $S_2$ instead
of $S_1$\cite{azuleneS1IC,azuleneS2S0IC,azuleneS2IC,azuleneS1S0conical,azulene_original}. As such, azulene is an ideal model for the study of IC phenomena.
For this reason, it is taken here as a showcase to reveal the performance of NAC-TDDFT in different flavors.
Since the AWF-based NAC-TDDFT is not at our disposal, we only report results by using the Hellmann-Feynman-like formula \eqref{HFey_rho}\cite{ChernyakMukamelNAC},
the EOM expressions \eqref{NAC_EOM_0I} and \eqref{NAC_EOM_IJ}\cite{lzdNACtheory,lzdNACimplementation},
as well as the TDPT ones \eqref{NAC_EOM_0I} and \eqref{RTee}\cite{lzdNACtheory,lzdNACimplementation}.
For the \emph{ee} fo-NACMEs described by Eq. \eqref{HFey_rho}, both the EOM rTDM $\tilde{\boldsymbol{\gamma}}^{IJ}$ in Eq. \eqref{ee-gammaEOM}
and the TDPT rTDM $\bar{\boldsymbol{\gamma}}^{IJ}$ in Eq. \eqref{TDPTrTDM}
were tested, with the corresponding results denoted as $g^{\xi,\mathrm{CBS(EOM)}}$ and $g^{\xi,\mathrm{CBS(TDPT)}}$, respectively. 
The $S_0$-$S_1$, $S_1$-$S_2$, and $T_1$-$T_2$ fo-NACMEs of azulene were computed with the BDF program package\cite{BDF1, BDF2, BDF3, BDFECC, BDFrev2020},
at the $S_1$, $S_2$ and $T_2$ equilibrium geometries
(optimized with B3LYP\cite{Becke93,B3LYP}/def2-SVP\cite{def2}), respectively. The response calculations employed the B3LYP collinear spin-flip exchange-correlation kernel,
for the noncollinear kernel\cite{NCOLL,RTDDFT1,RTDDFT2} has not yet been interfaced with the NAC-TDDFT module in BDF.

\begin{figure}
	\begin{tabular}{lcccc} 
		& def2-SV(P) & def2-SVP & def2-TZVP & def2-TZVPP \\
		
		$g_{S_0S_1}^{\xi, \mathrm{EOM}}$&
		\resizebox{0.15\textwidth}{!}{\includegraphics[angle=90, trim=600 0 600 0, clip]{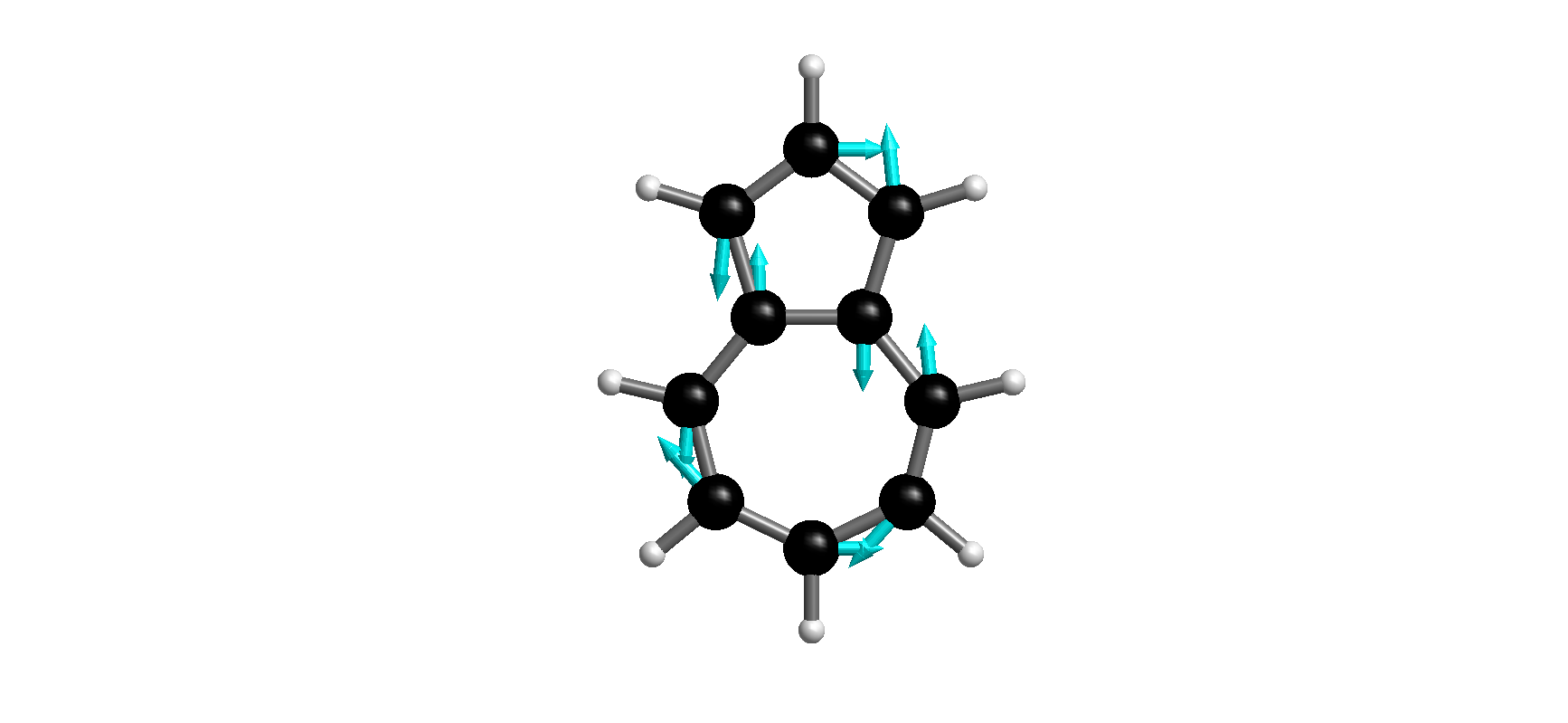}} &
		\resizebox{0.15\textwidth}{!}{\includegraphics[angle=90, trim=600 0 600 0, clip]{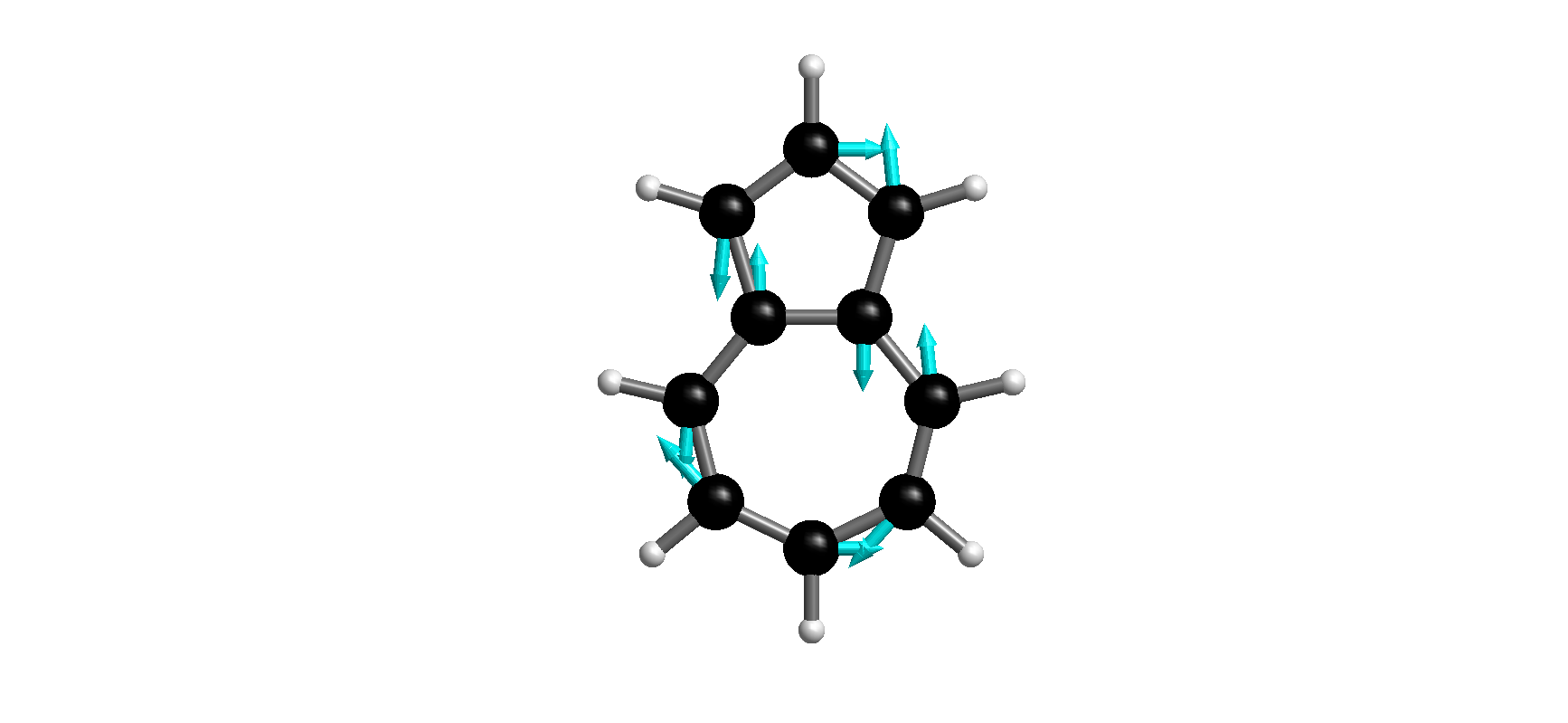}} &
		\resizebox{0.15\textwidth}{!}{\includegraphics[angle=90, trim=600 0 600 0, clip]{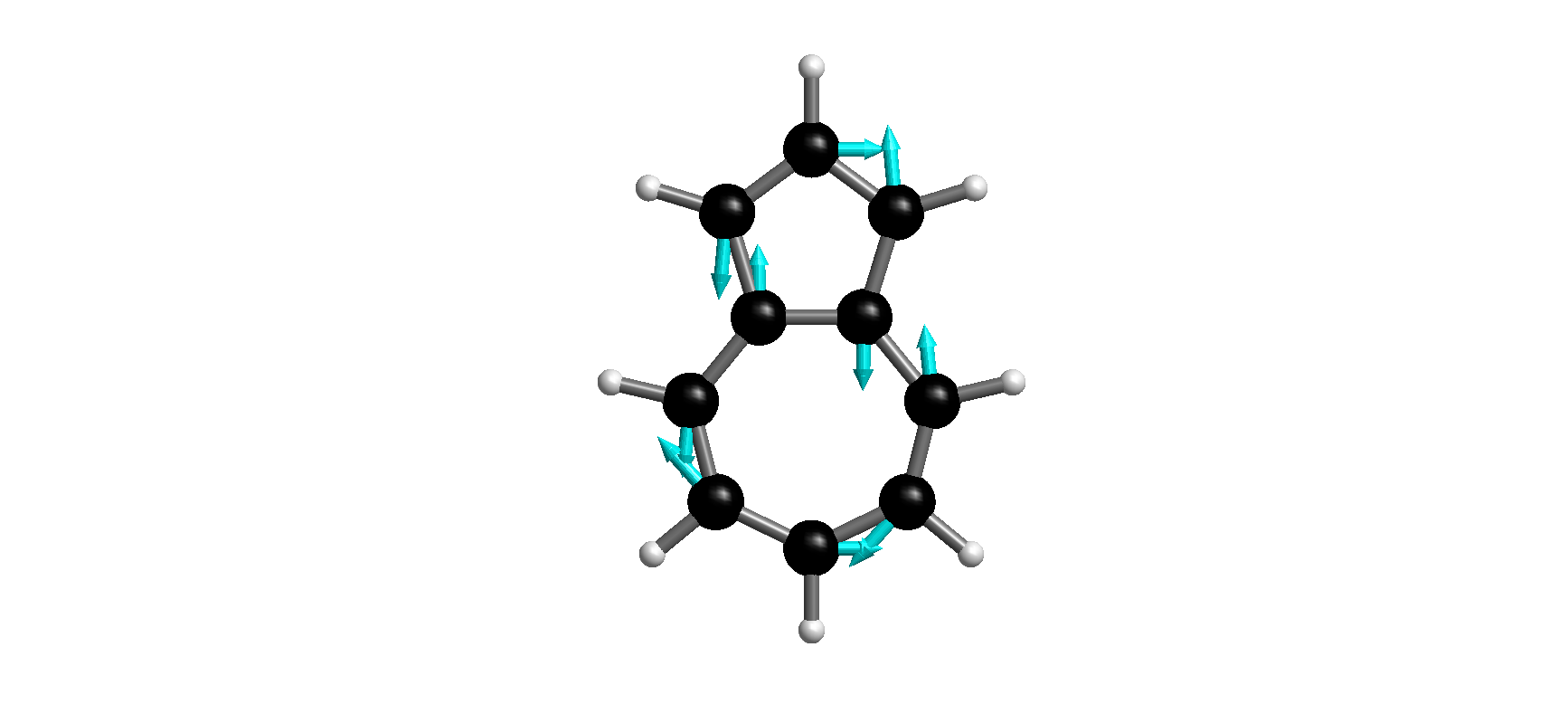}} &
		\resizebox{0.15\textwidth}{!}{\includegraphics[angle=90, trim=600 0 600 0, clip]{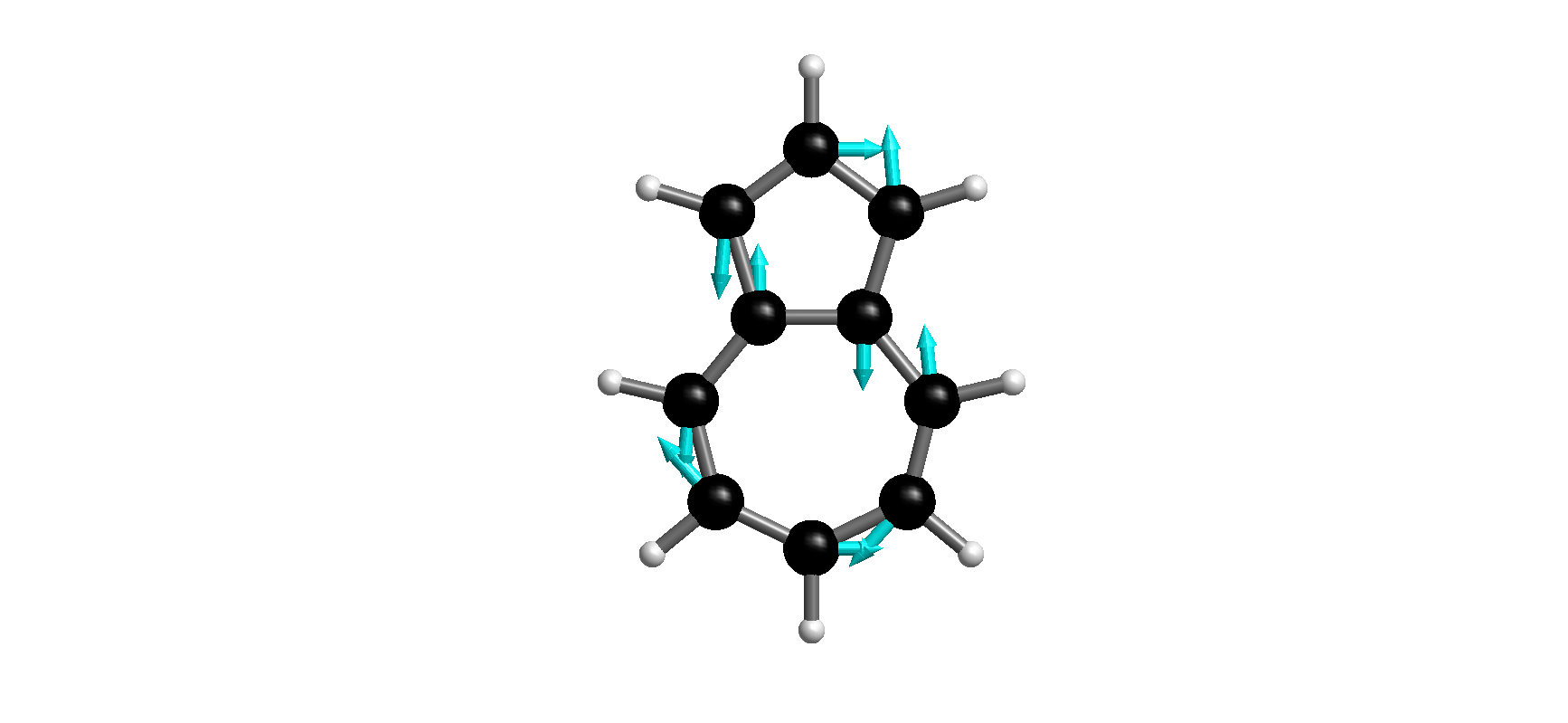}} \\
		
		$g_{S_0S_1}^{\xi, \mathrm{CBS(EOM)}}$&
		\resizebox{0.15\textwidth}{!}{\includegraphics[angle=90, trim=600 0 600 0, clip]{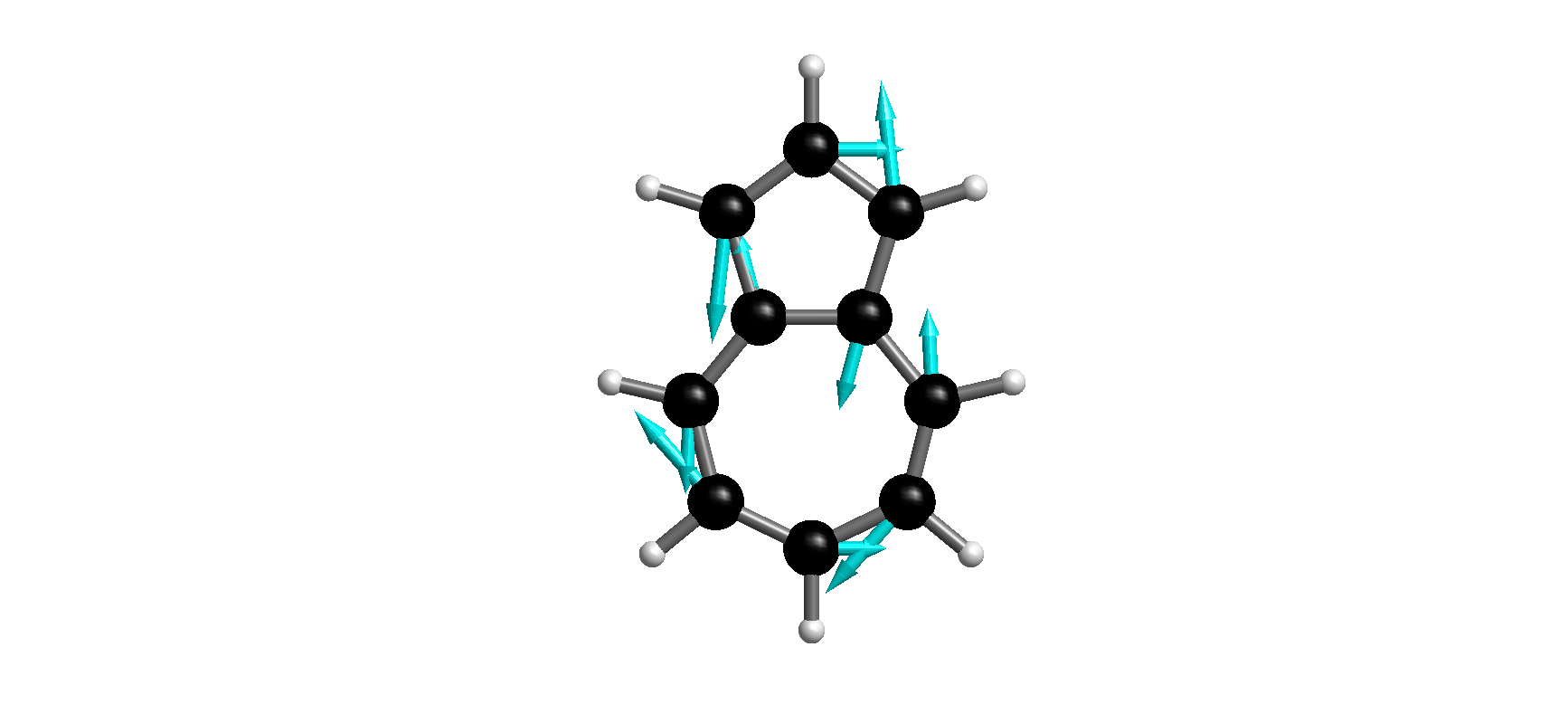}} &
		\resizebox{0.15\textwidth}{!}{\includegraphics[angle=90, trim=600 0 600 0, clip]{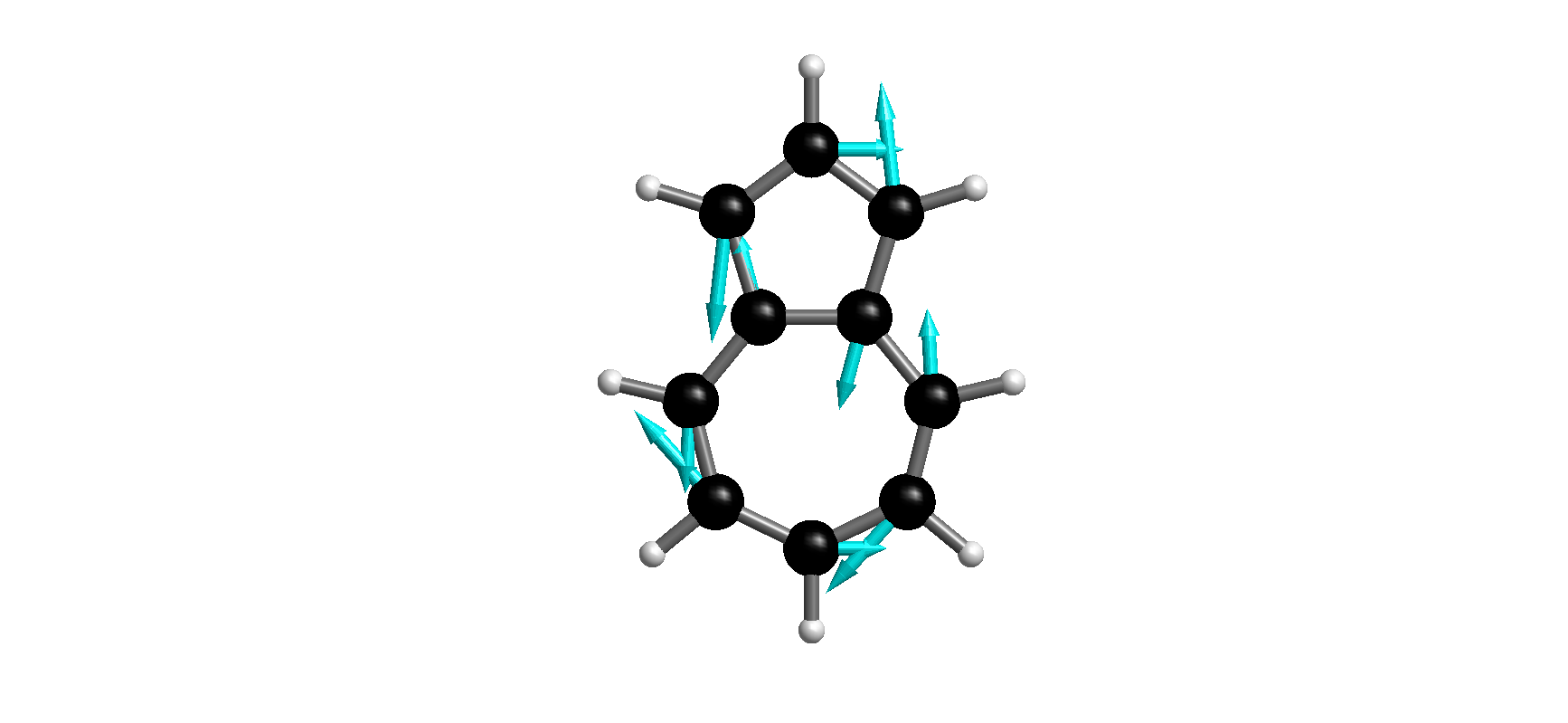}} &
		\resizebox{0.15\textwidth}{!}{\includegraphics[angle=90, trim=600 0 600 0, clip]{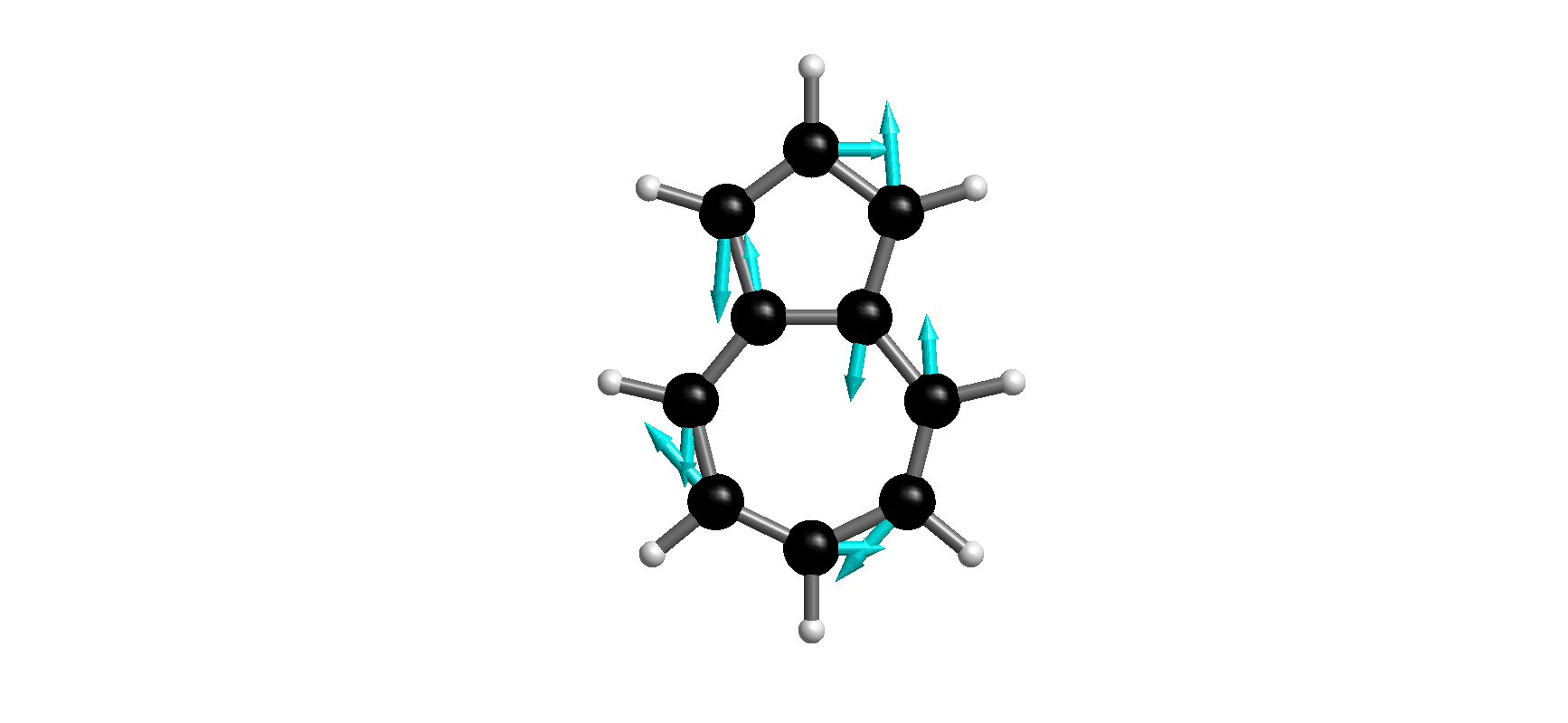}} &
		\resizebox{0.15\textwidth}{!}{\includegraphics[angle=90, trim=600 0 600 0, clip]{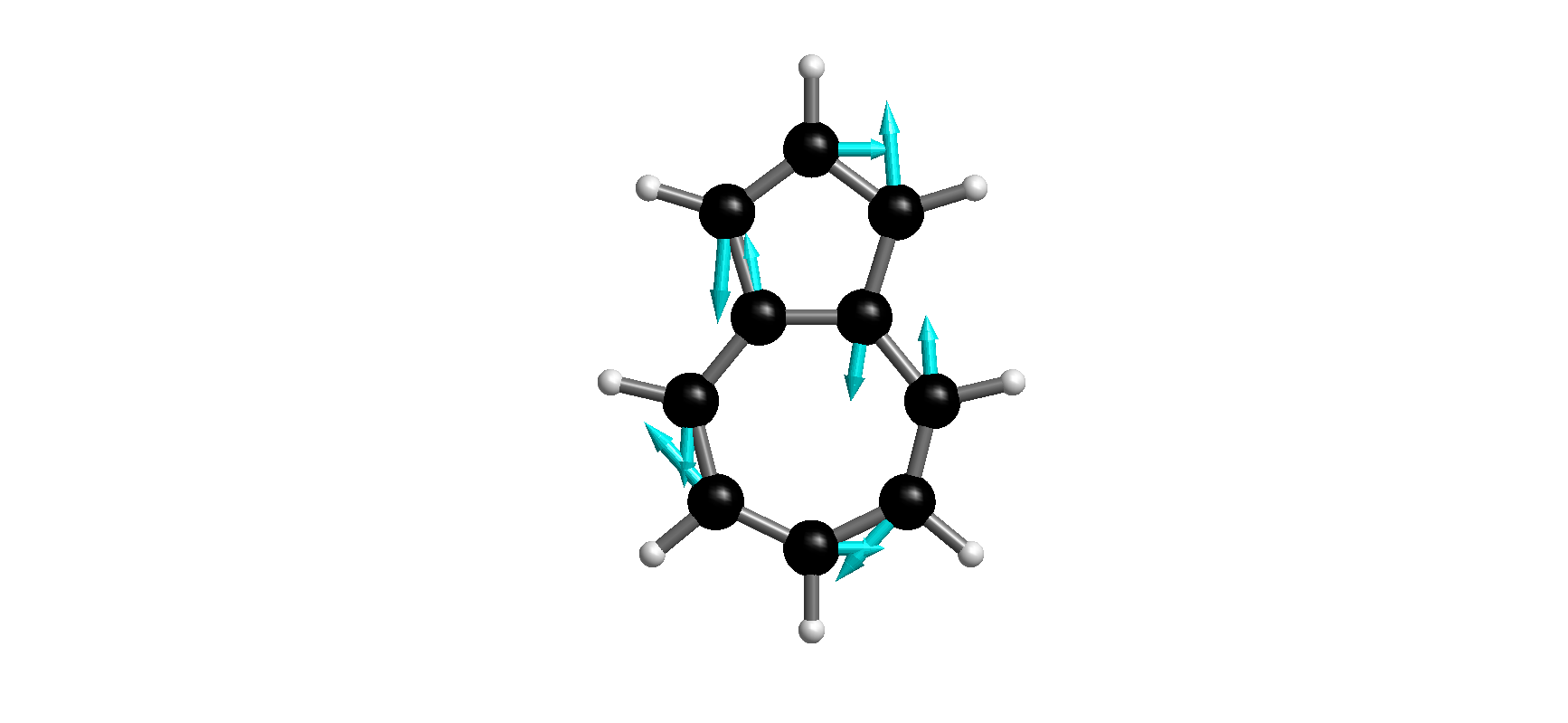}} \\
		
		$g_{S_1S_2}^{\xi, \mathrm{EOM}}$&
		\resizebox{0.15\textwidth}{!}{\includegraphics[angle=90, trim=600 0 600 0, clip]{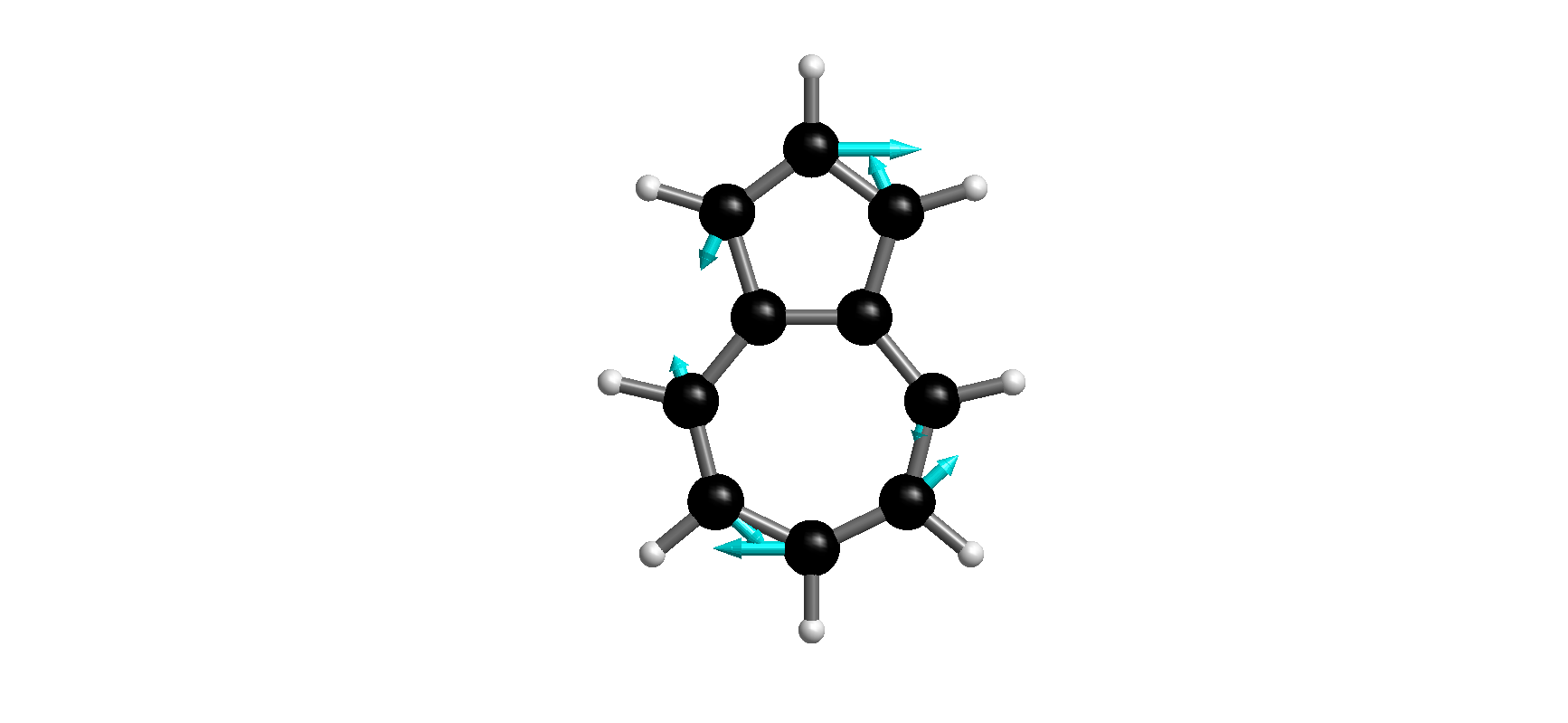}} &
		\resizebox{0.15\textwidth}{!}{\includegraphics[angle=90, trim=600 0 600 0, clip]{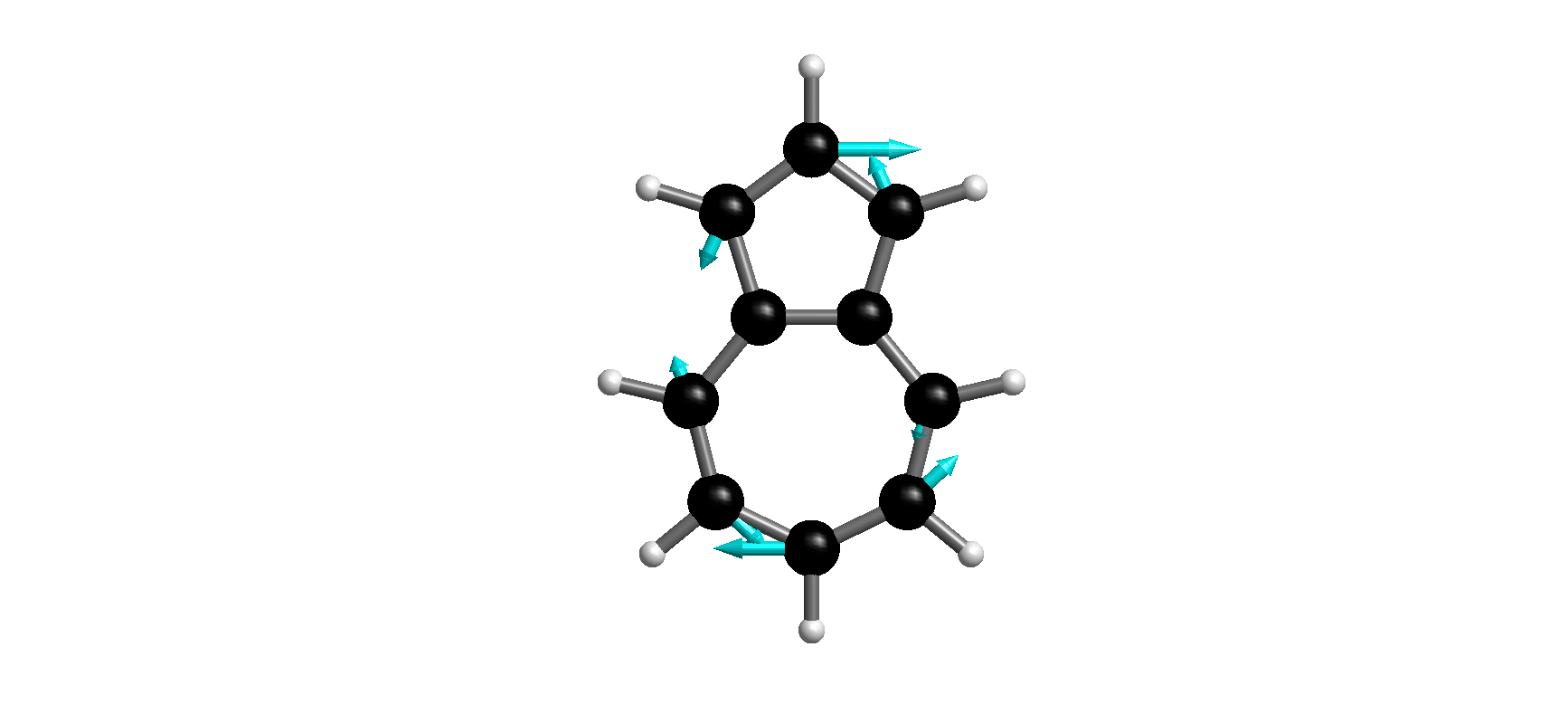}} &
		\resizebox{0.15\textwidth}{!}{\includegraphics[angle=90, trim=600 0 600 0, clip]{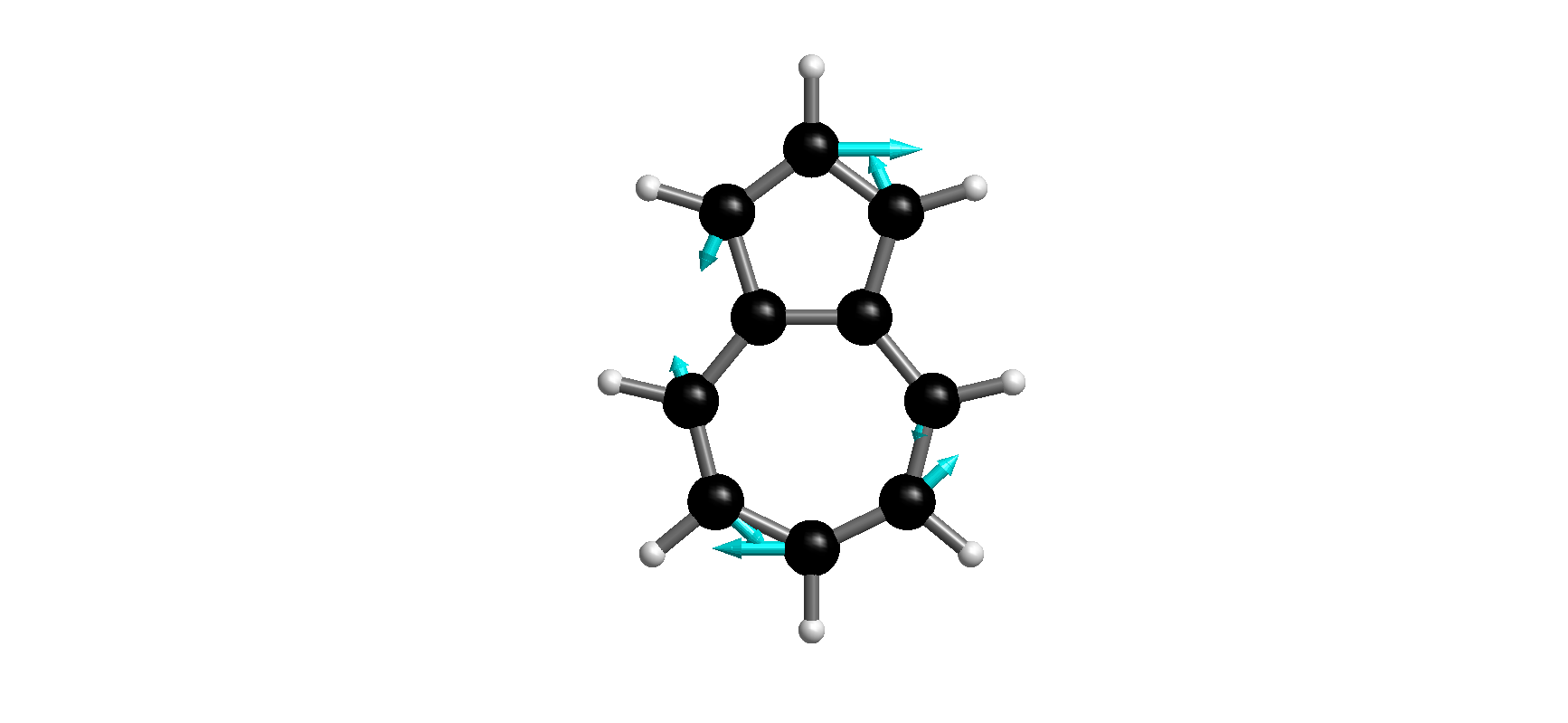}} &
		\resizebox{0.15\textwidth}{!}{\includegraphics[angle=90, trim=600 0 600 0, clip]{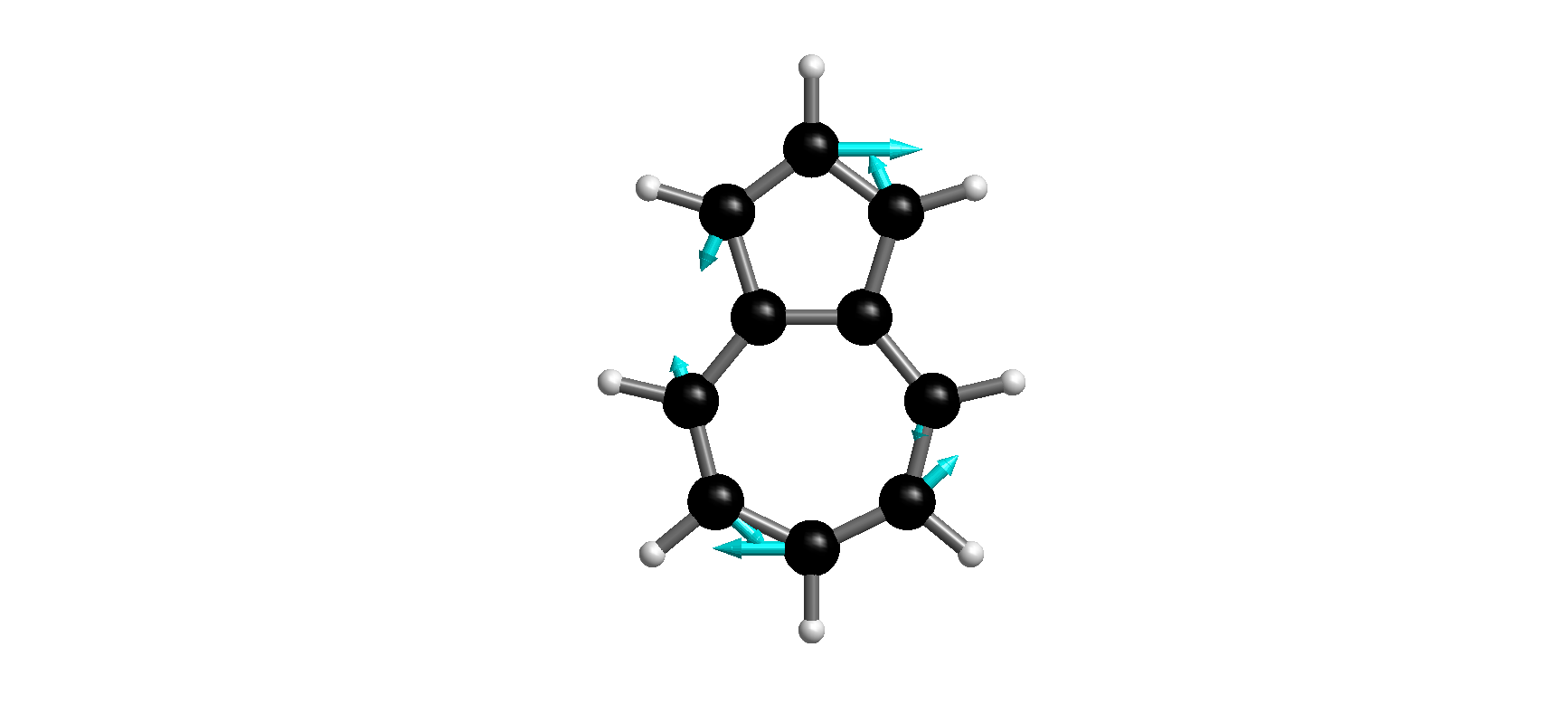}} \\
		
		$g_{S_1S_2}^{\xi, \mathrm{TDPT}}$&
		\resizebox{0.15\textwidth}{!}{\includegraphics[angle=90, trim=600 0 600 0, clip]{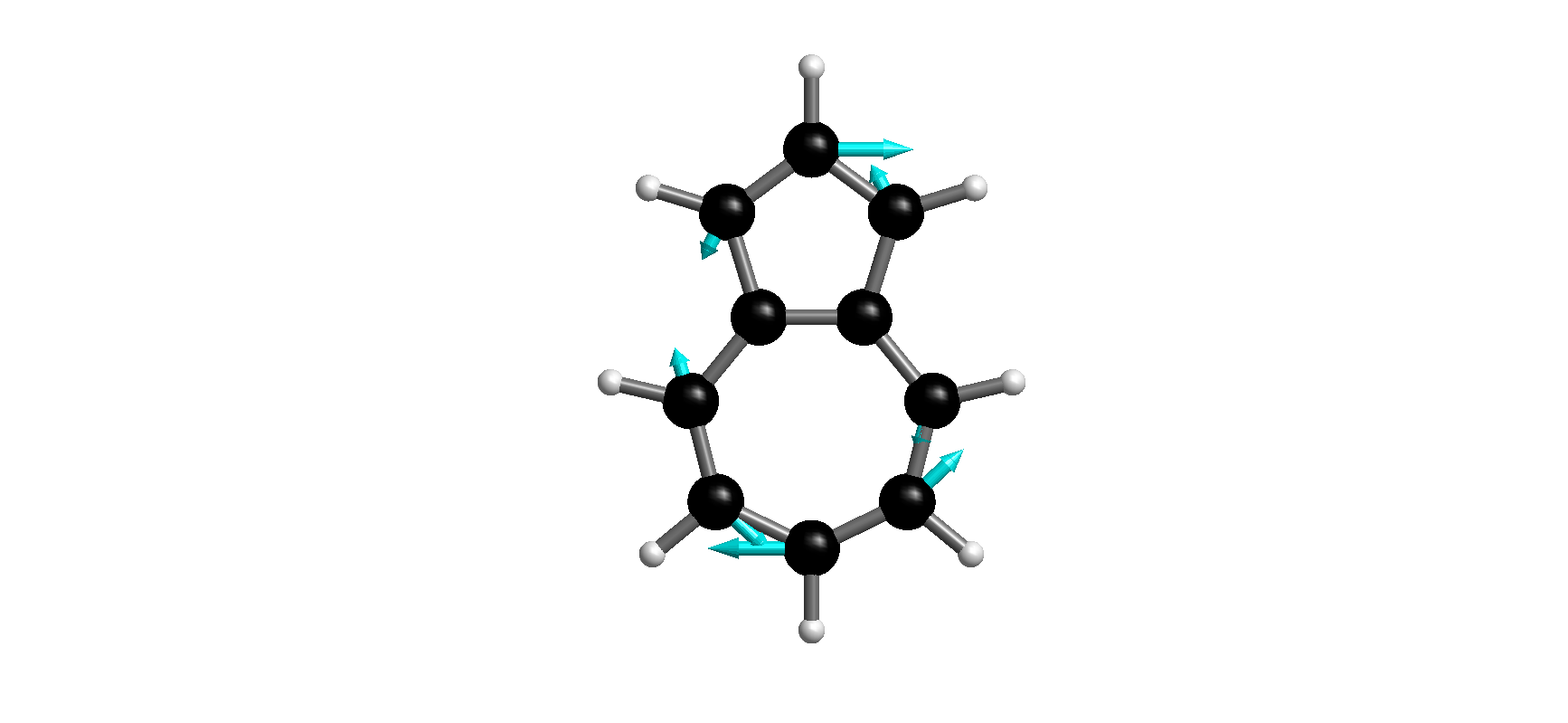}} &
		\resizebox{0.15\textwidth}{!}{\includegraphics[angle=90, trim=600 0 600 0, clip]{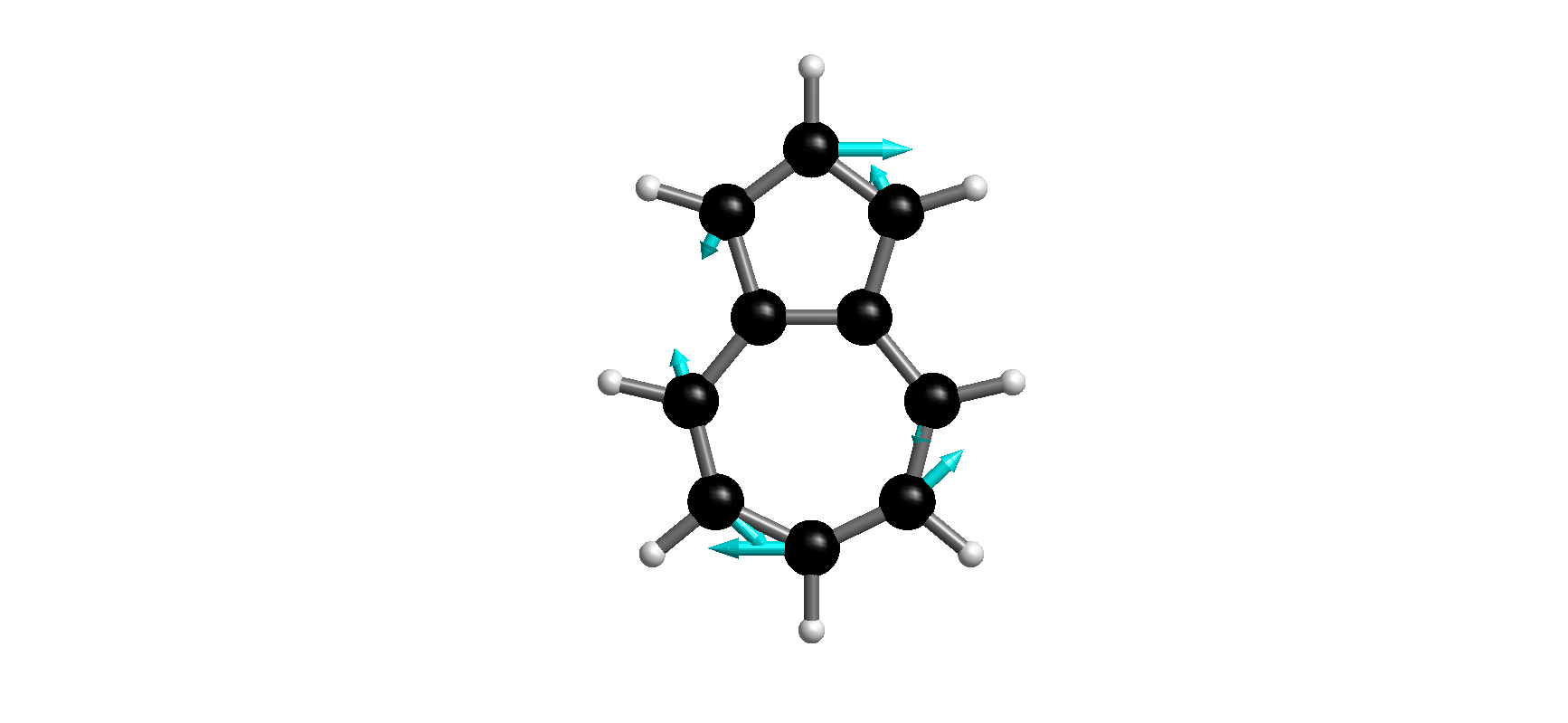}} &
		\resizebox{0.15\textwidth}{!}{\includegraphics[angle=90, trim=600 0 600 0, clip]{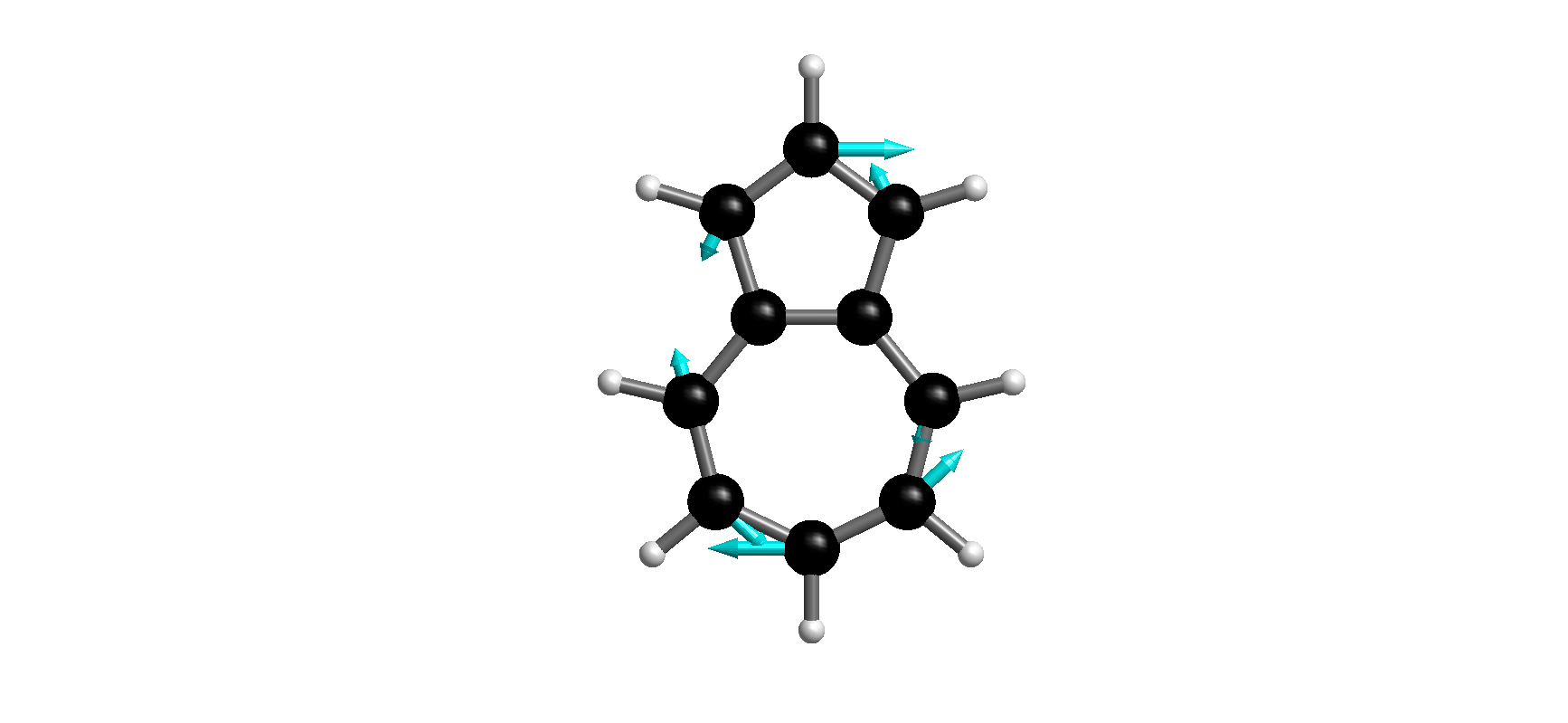}} &
		\resizebox{0.15\textwidth}{!}{\includegraphics[angle=90, trim=600 0 600 0, clip]{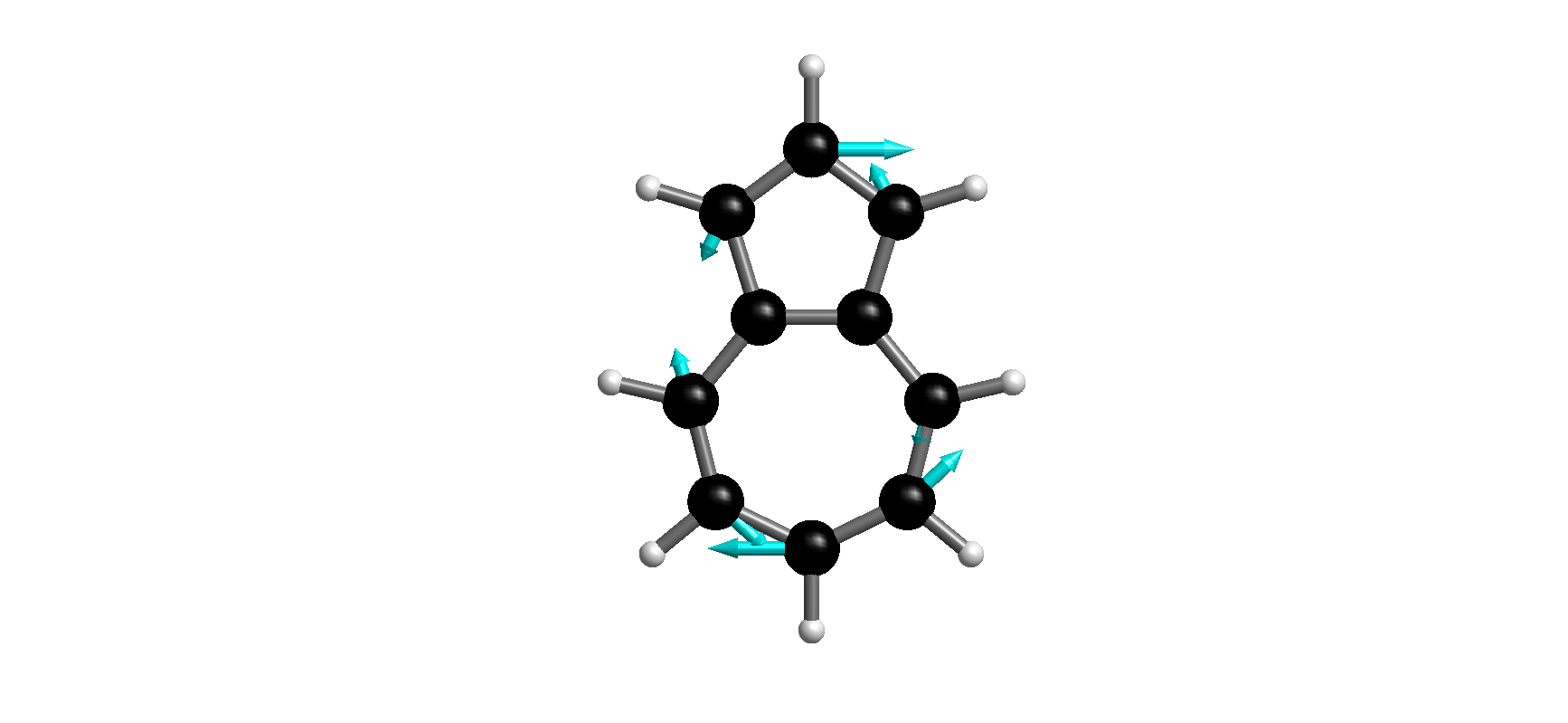}} \\
		
		$g_{S_1S_2}^{\xi, \mathrm{CBS(EOM)}}$&
		\resizebox{0.15\textwidth}{!}{\includegraphics[angle=90, trim=550 0 500 0, clip]{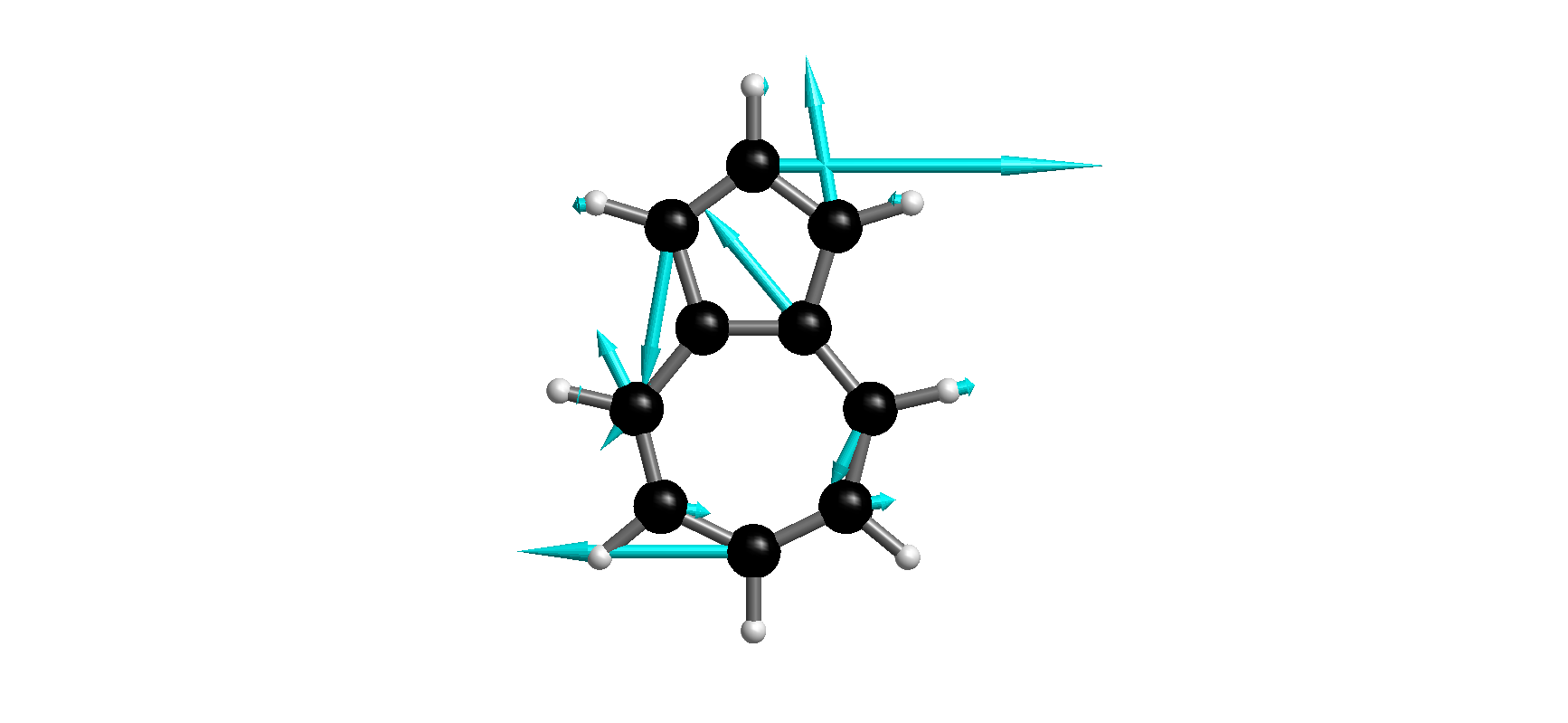}} &
		\resizebox{0.15\textwidth}{!}{\includegraphics[angle=90, trim=550 0 500 0, clip]{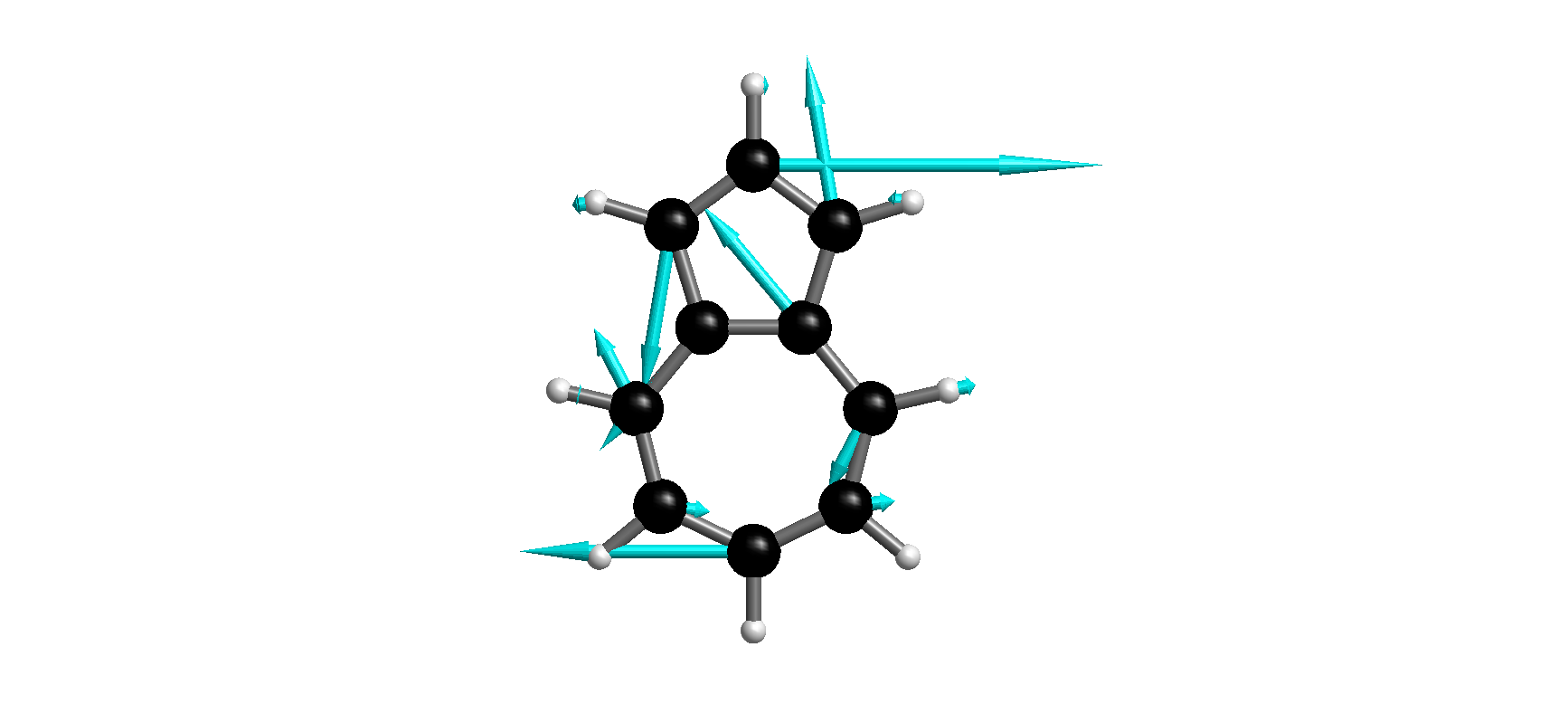}} &
		\resizebox{0.15\textwidth}{!}{\includegraphics[angle=90, trim=550 0 500 0, clip]{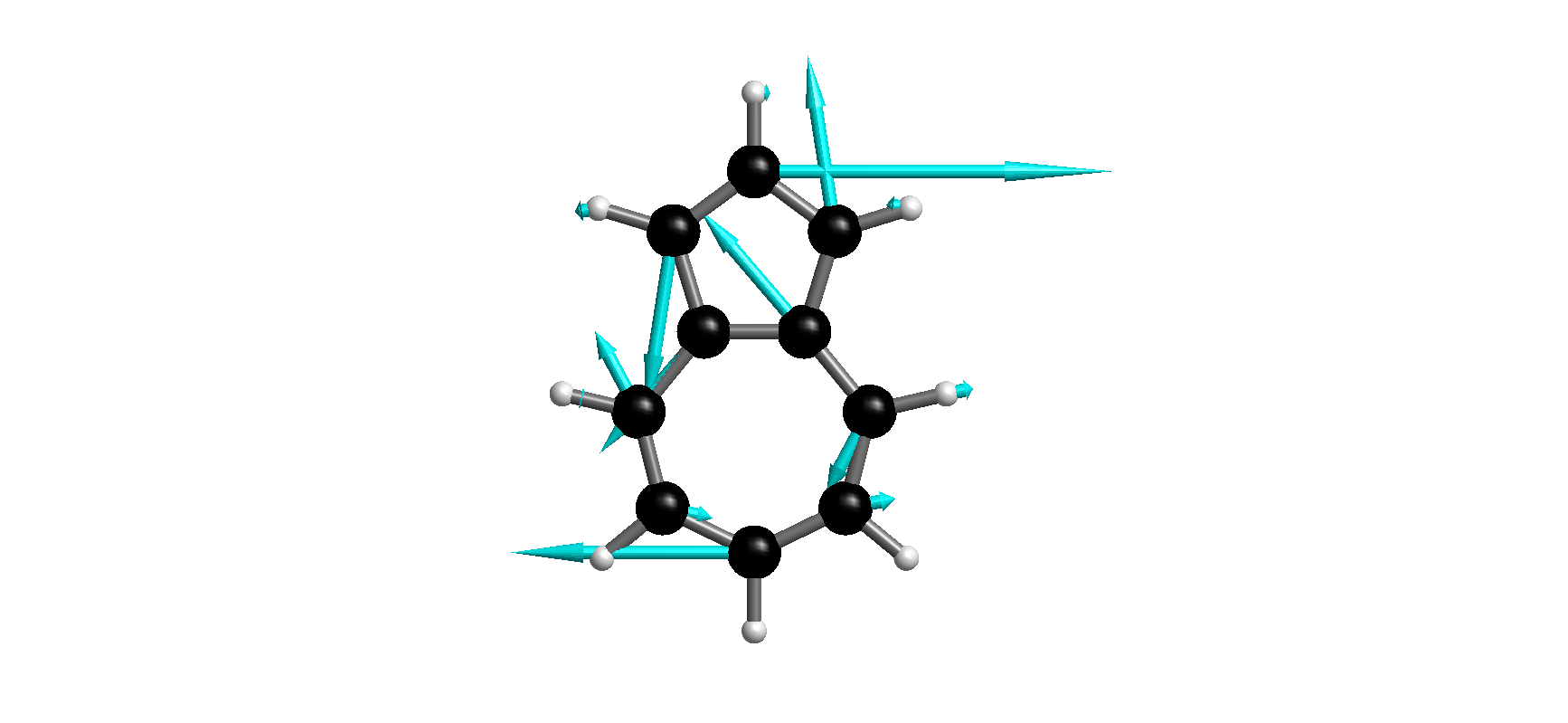}} &
		\resizebox{0.15\textwidth}{!}{\includegraphics[angle=90, trim=550 0 500 0, clip]{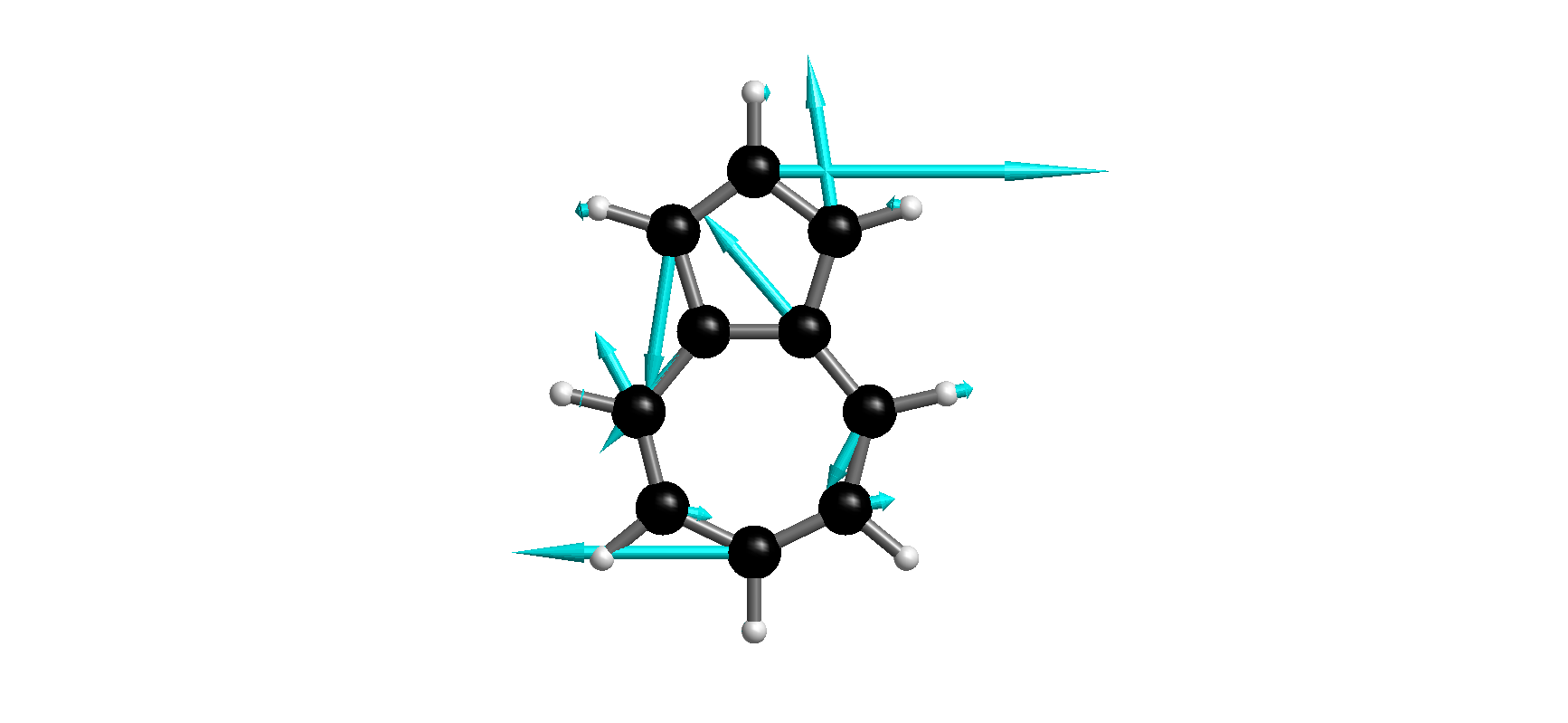}} \\
		
		$g_{S_1S_2}^{\xi, \mathrm{CBS(TDPT)}}$&
		\resizebox{0.15\textwidth}{!}{\includegraphics[angle=90, trim=600 0 500 0, clip]{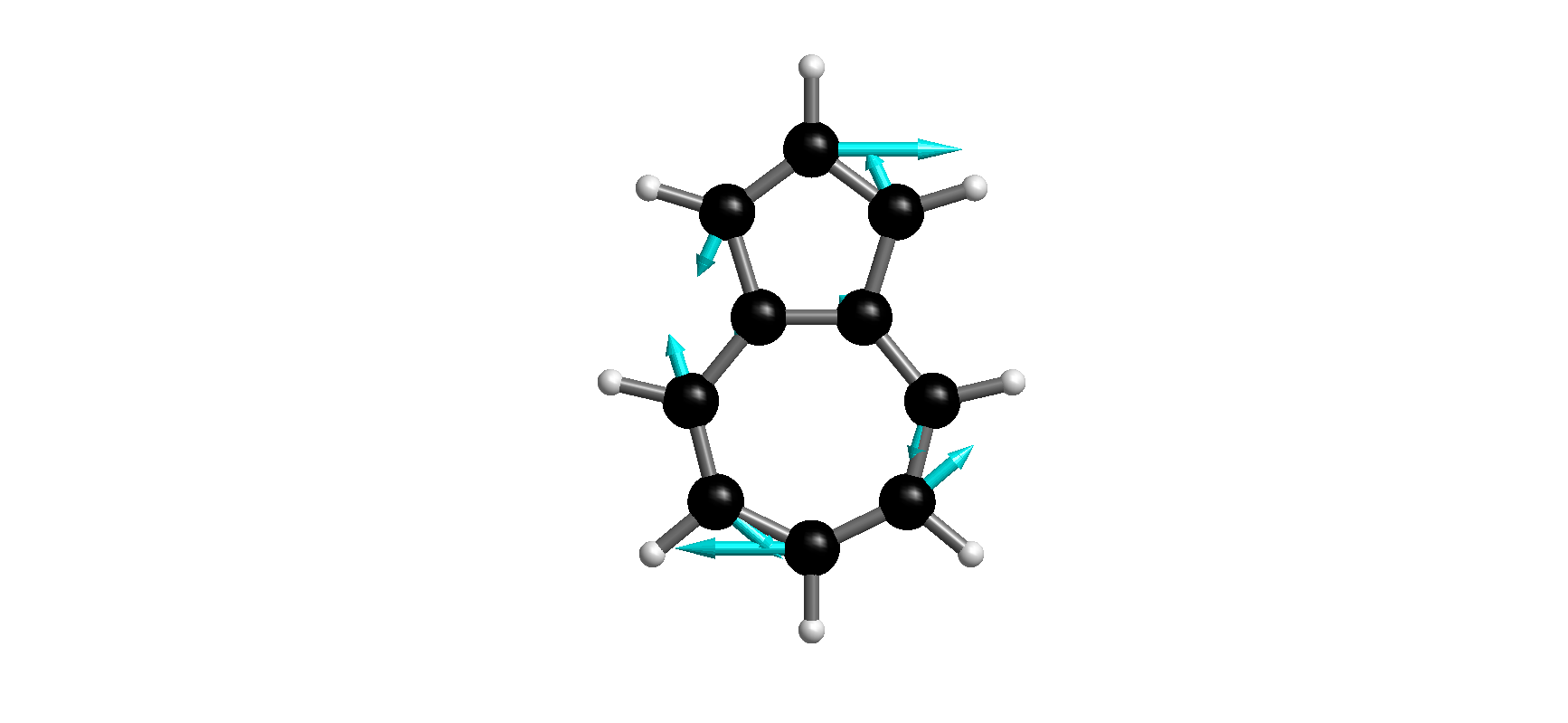}} &
		\resizebox{0.15\textwidth}{!}{\includegraphics[angle=90, trim=600 0 500 0, clip]{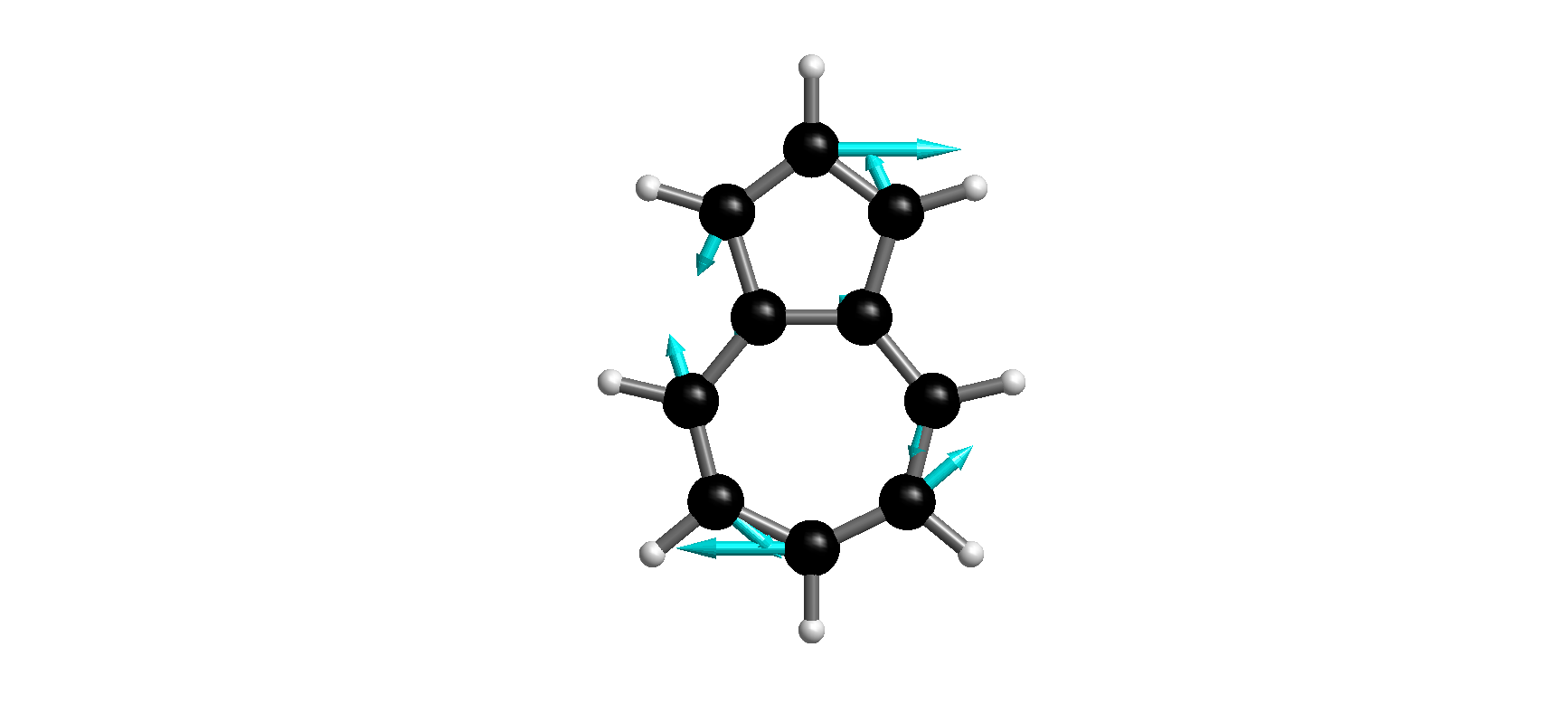}} &
		\resizebox{0.15\textwidth}{!}{\includegraphics[angle=90, trim=600 0 500 0, clip]{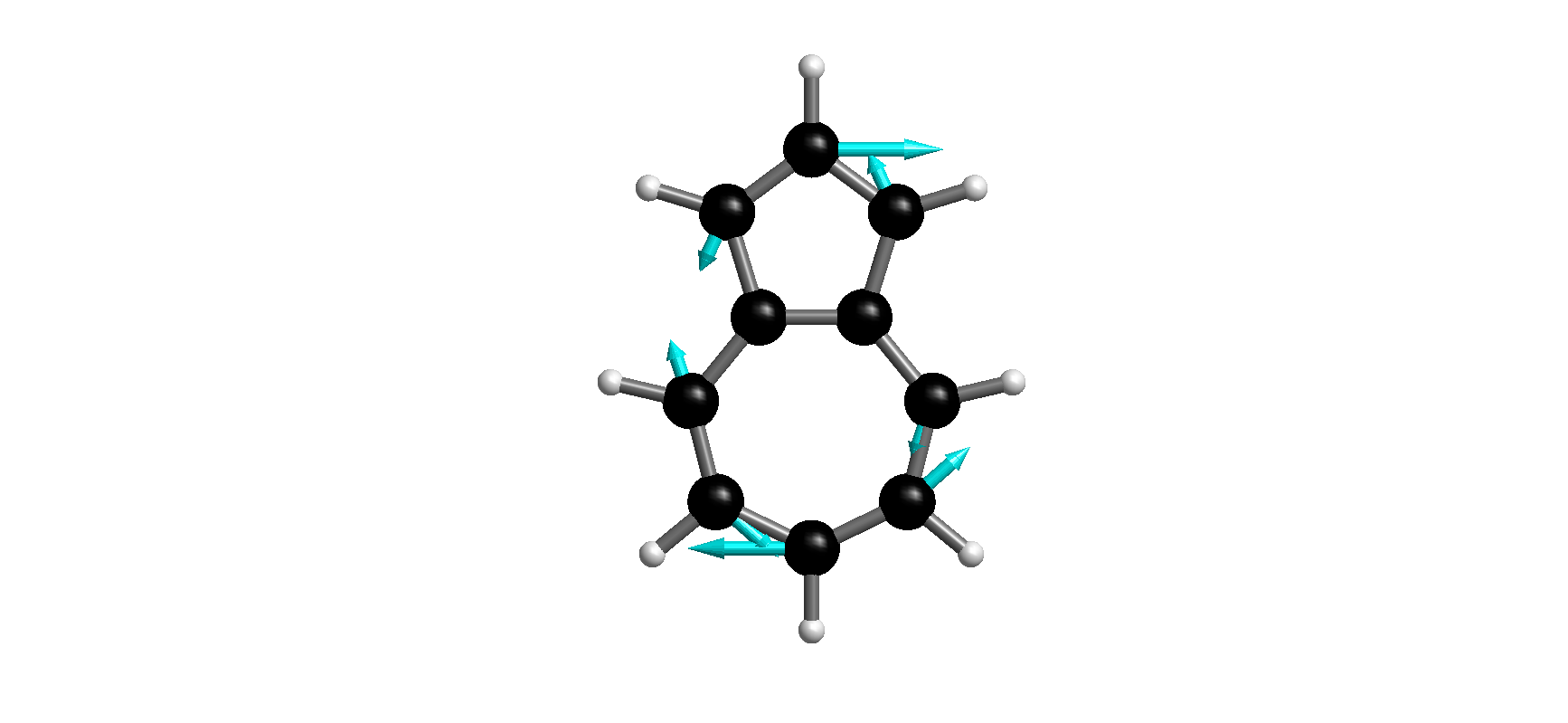}} &
		\resizebox{0.15\textwidth}{!}{\includegraphics[angle=90, trim=600 0 500 0, clip]{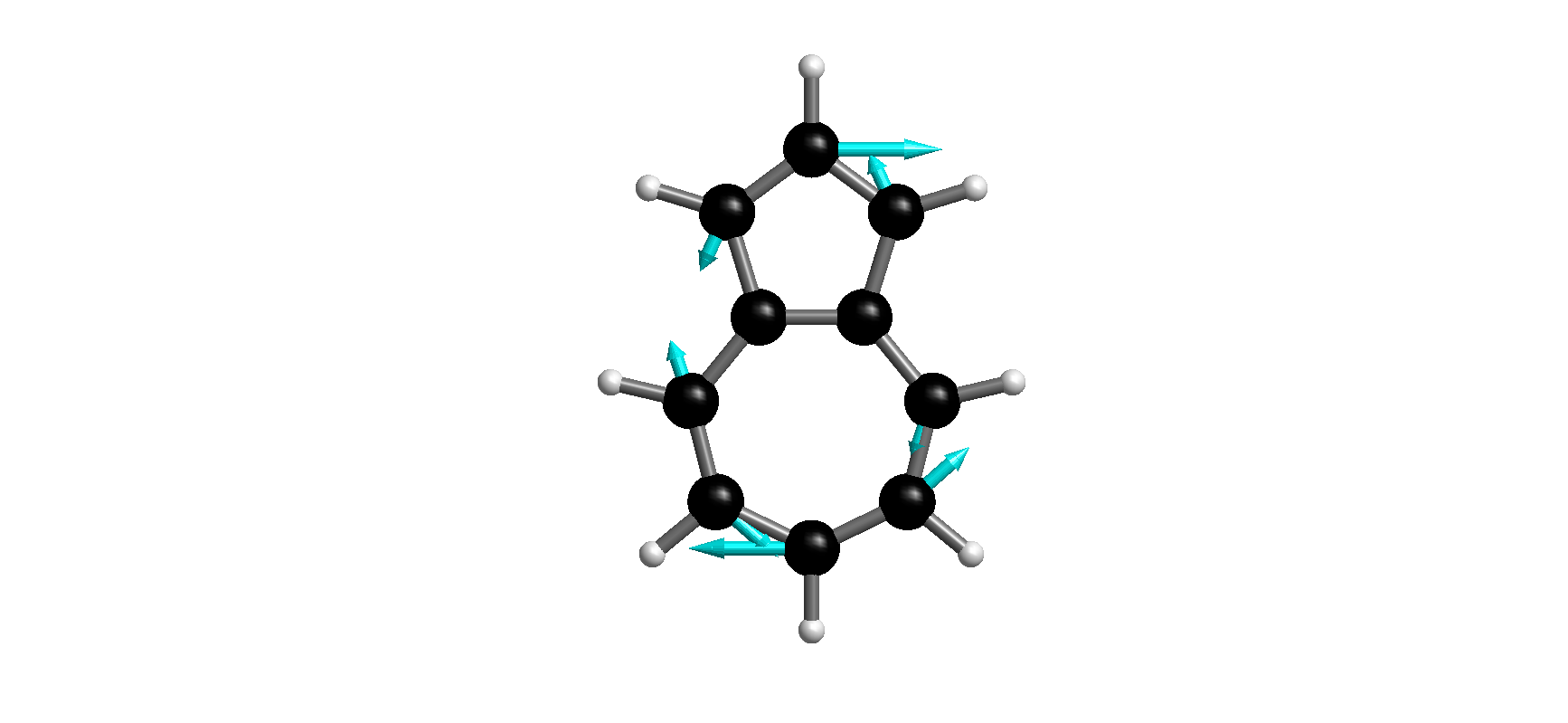}} \\
		
		$g_{T_1T_2}^{\xi, \mathrm{EOM}}$&
		\resizebox{0.15\textwidth}{!}{\includegraphics[angle=90, trim=600 0 600 0, clip]{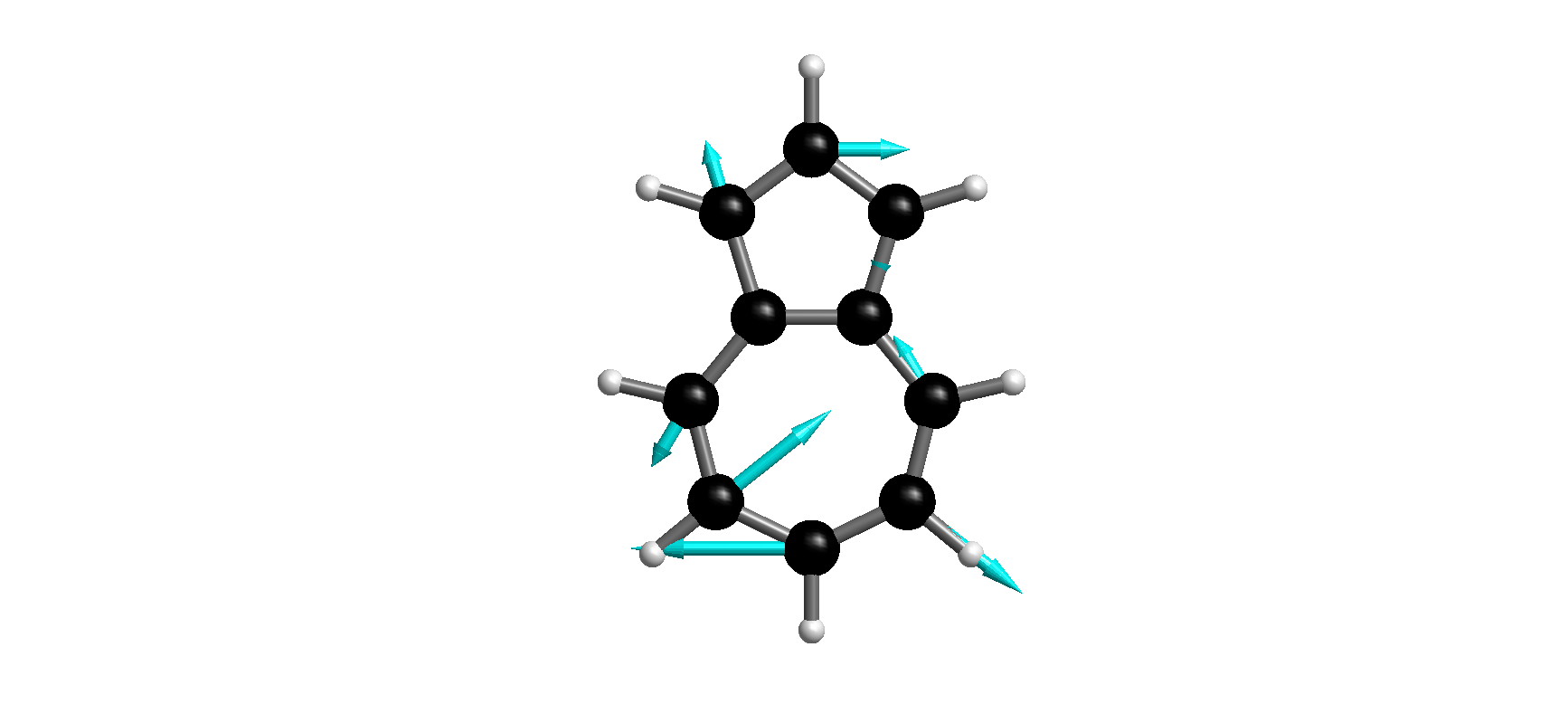}} &
		\resizebox{0.15\textwidth}{!}{\includegraphics[angle=90, trim=600 0 600 0, clip]{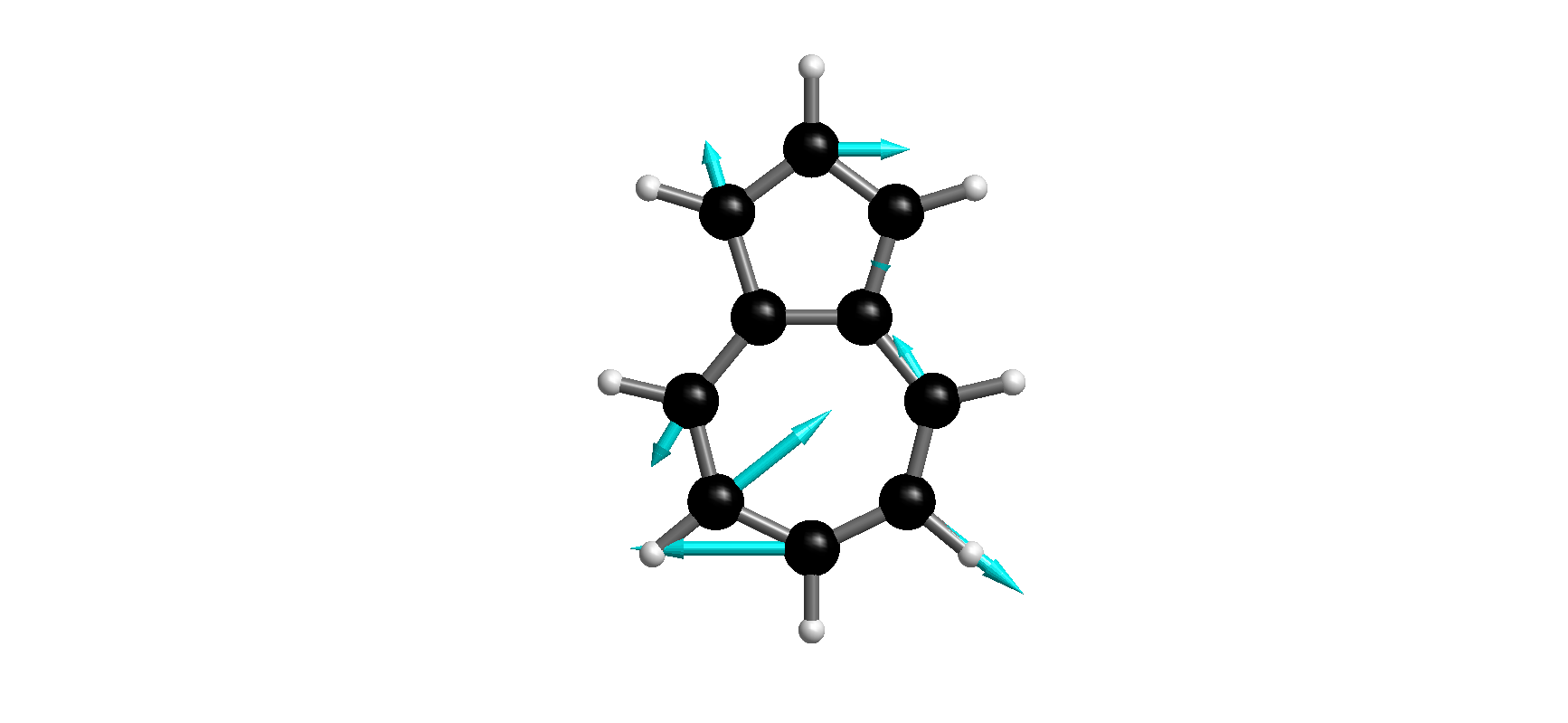}} &
		\resizebox{0.15\textwidth}{!}{\includegraphics[angle=90, trim=600 0 600 0, clip]{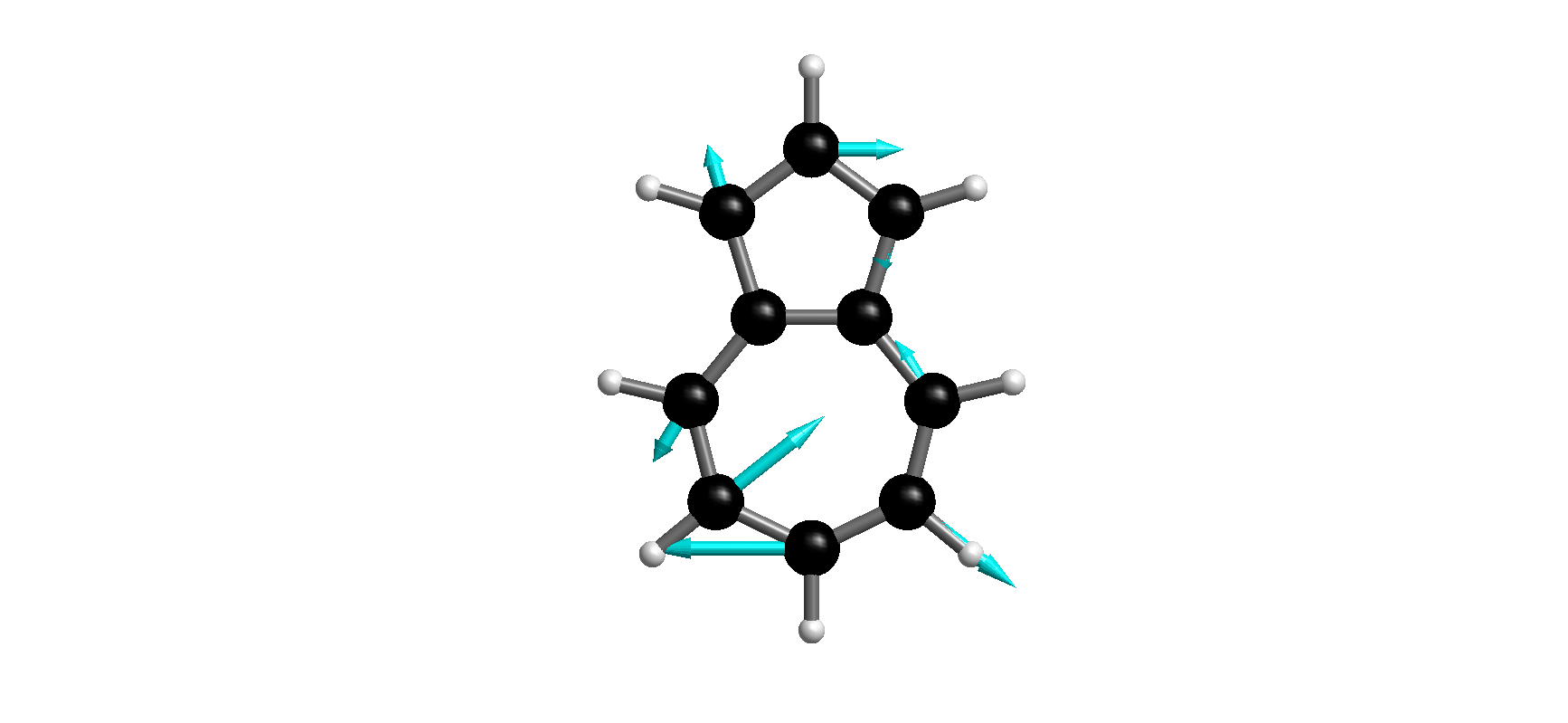}} &
		\resizebox{0.15\textwidth}{!}{\includegraphics[angle=90, trim=600 0 600 0, clip]{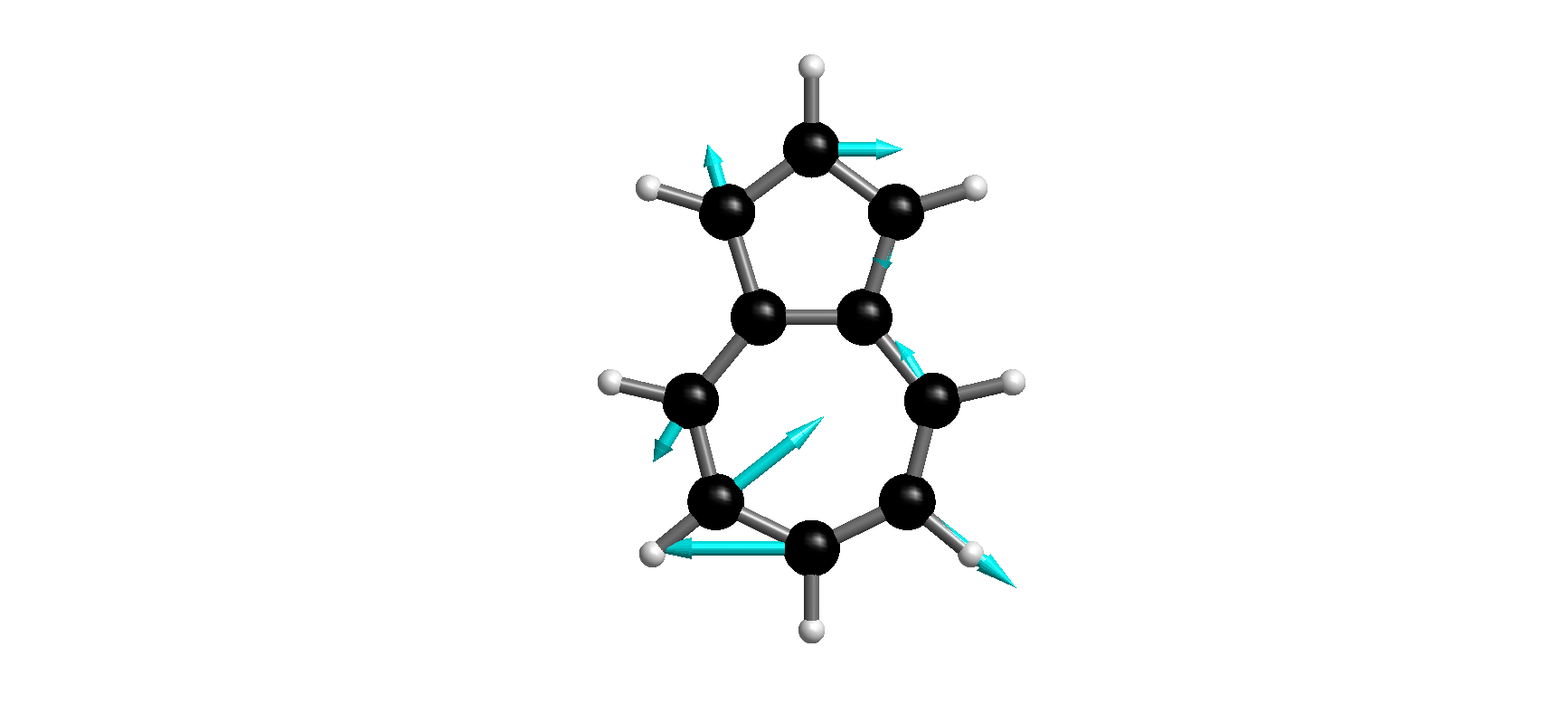}} \\
		
		$g_{T_1T_2}^{\xi, \mathrm{TDPT}}$&
		\resizebox{0.15\textwidth}{!}{\includegraphics[angle=90, trim=600 0 600 0, clip]{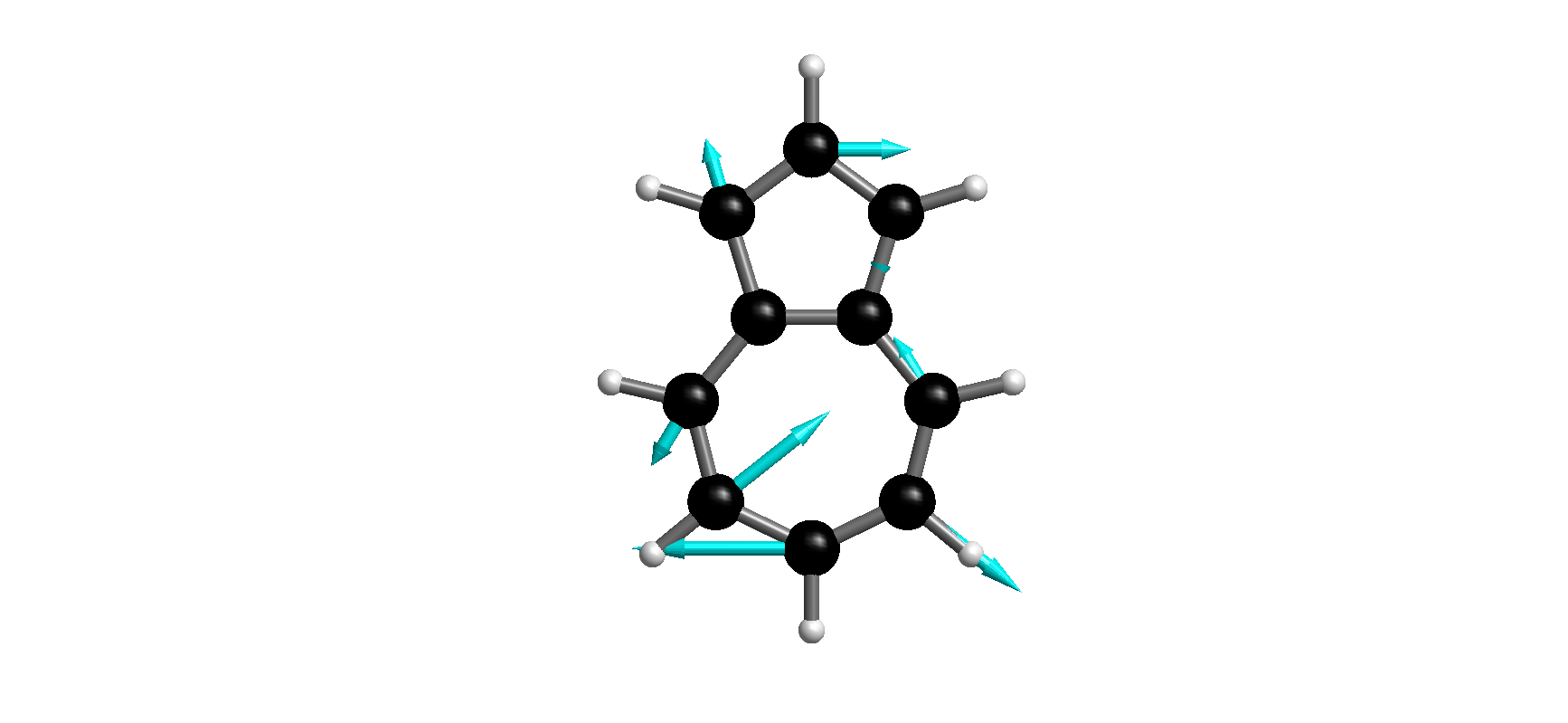}} &
		\resizebox{0.15\textwidth}{!}{\includegraphics[angle=90, trim=600 0 600 0, clip]{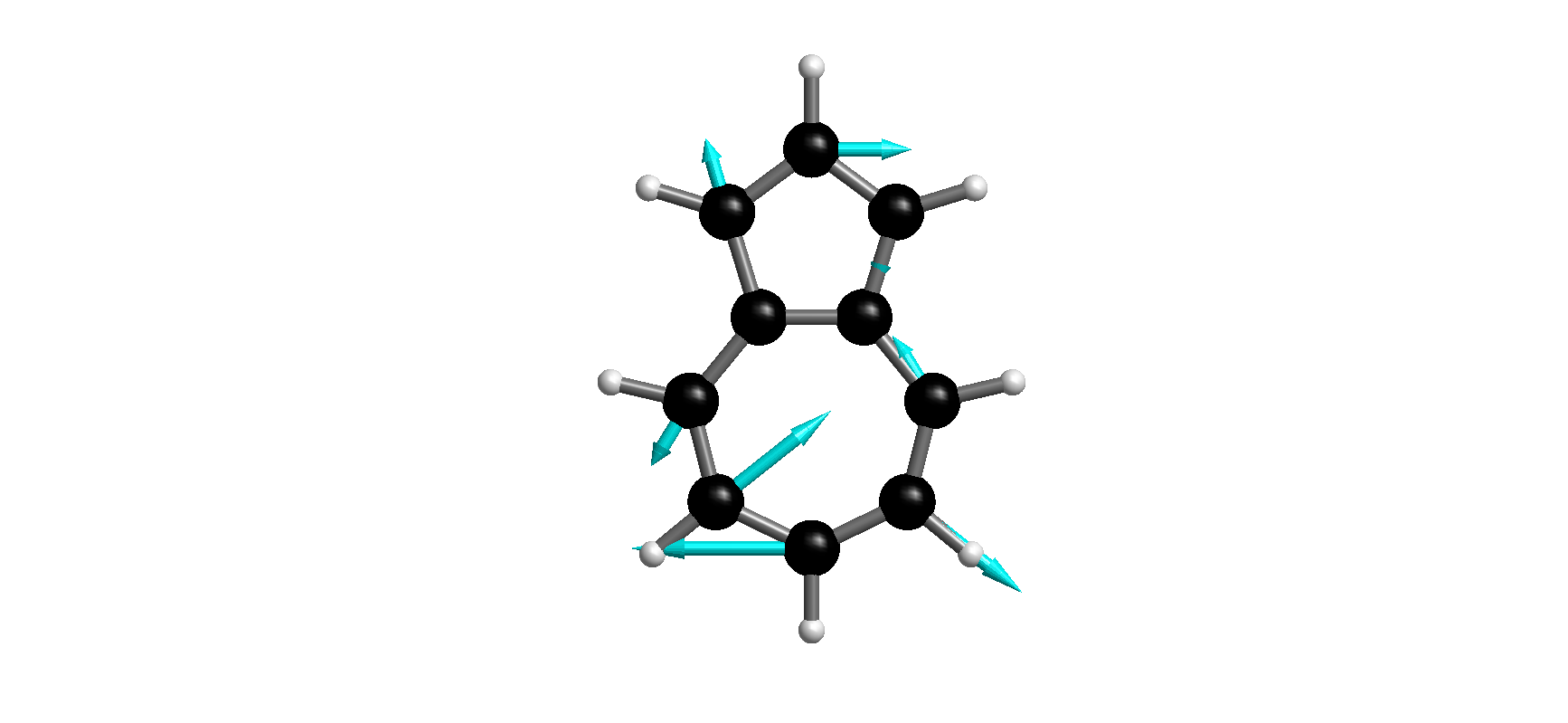}} &
		\resizebox{0.15\textwidth}{!}{\includegraphics[angle=90, trim=600 0 600 0, clip]{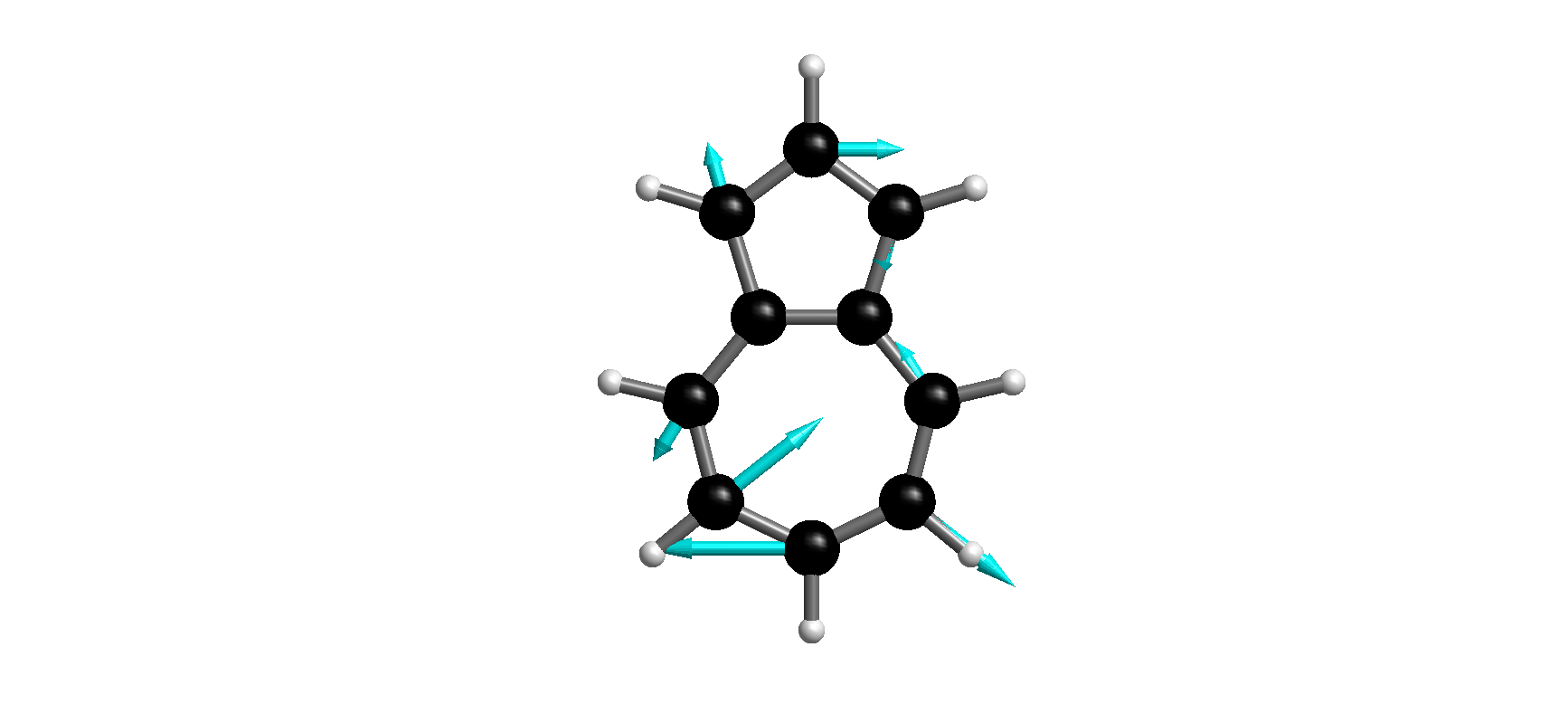}} &
		\resizebox{0.15\textwidth}{!}{\includegraphics[angle=90, trim=600 0 600 0, clip]{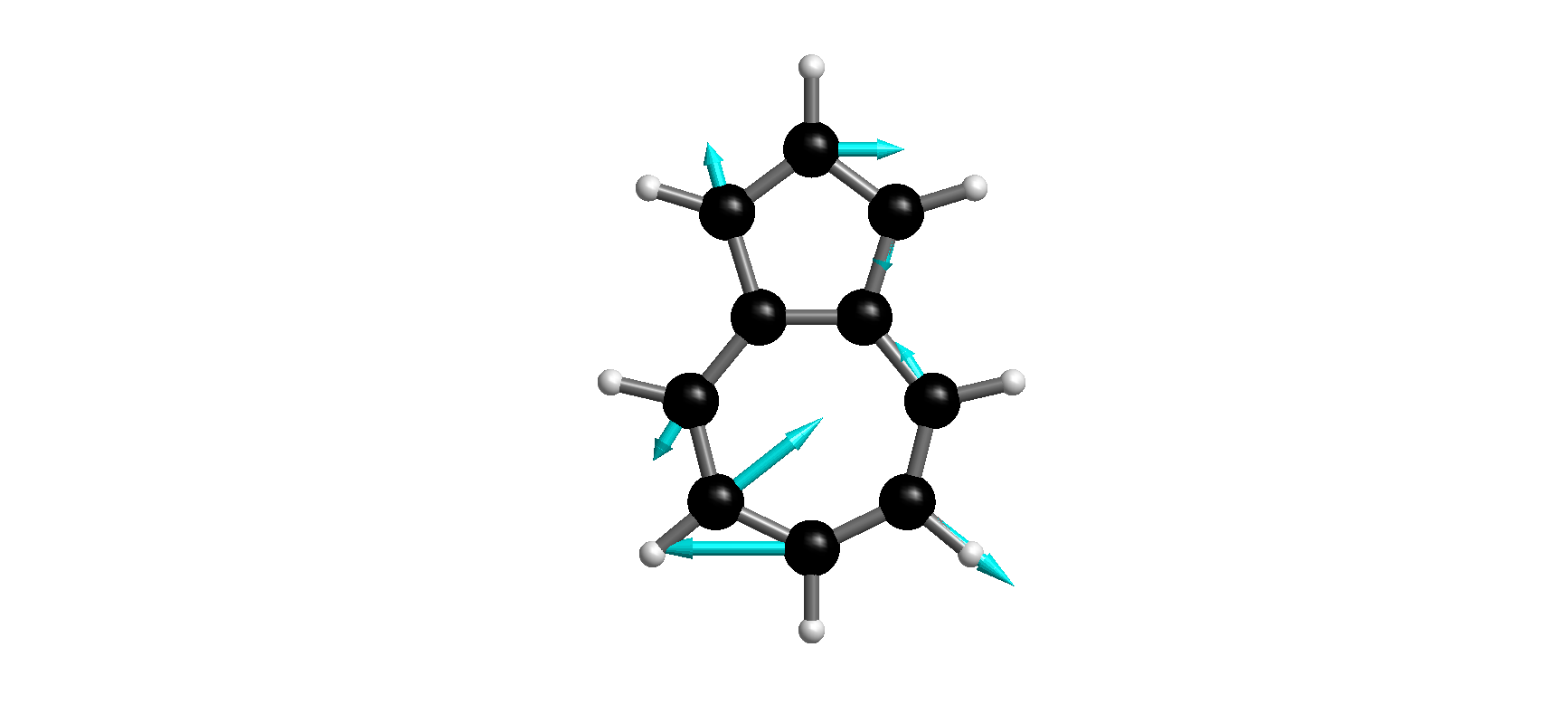}} \\
		
		$g_{T_1T_2}^{\xi, \mathrm{CBS(EOM)}}$&
		\resizebox{0.15\textwidth}{!}{\includegraphics[angle=90, trim=600 0 600 0, clip]{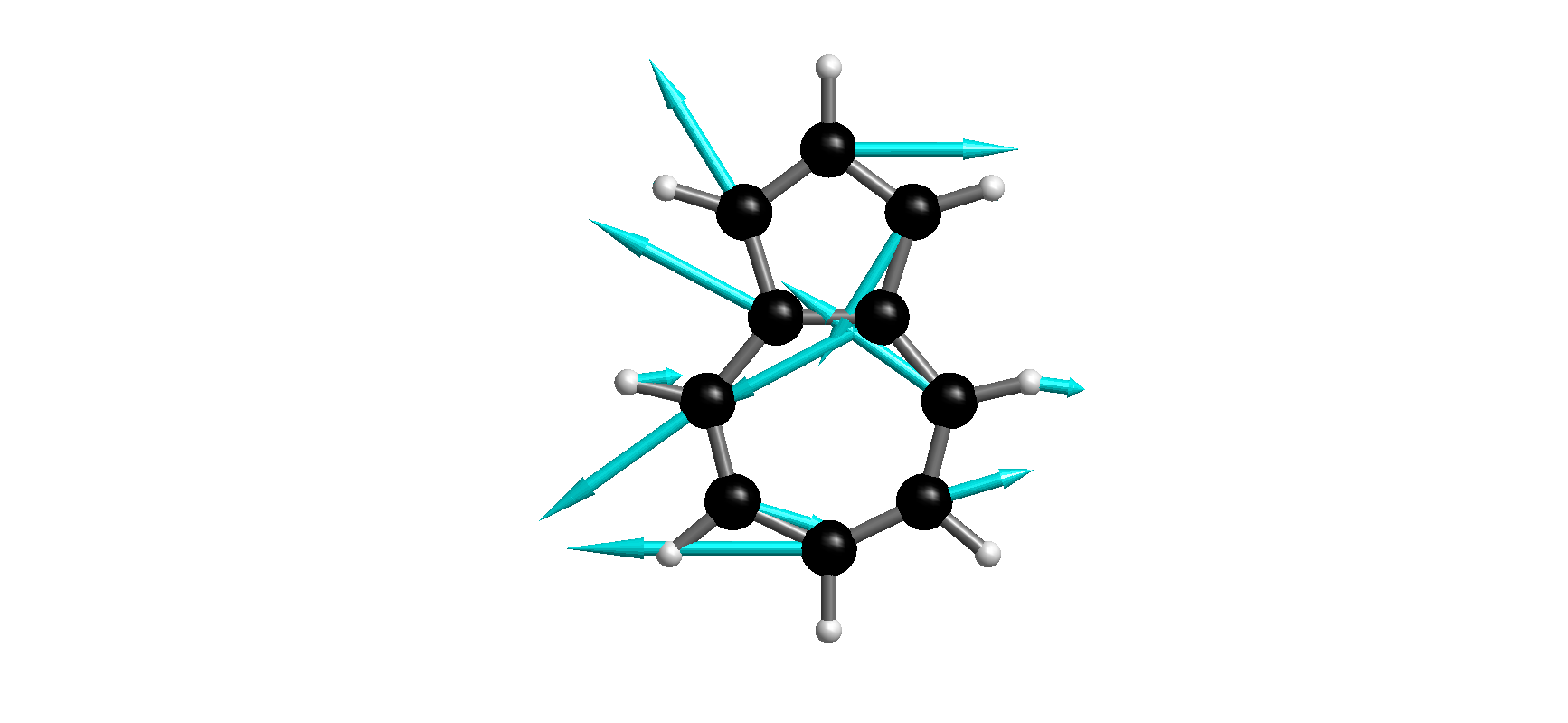}} &
		\resizebox{0.15\textwidth}{!}{\includegraphics[angle=90, trim=600 0 600 0, clip]{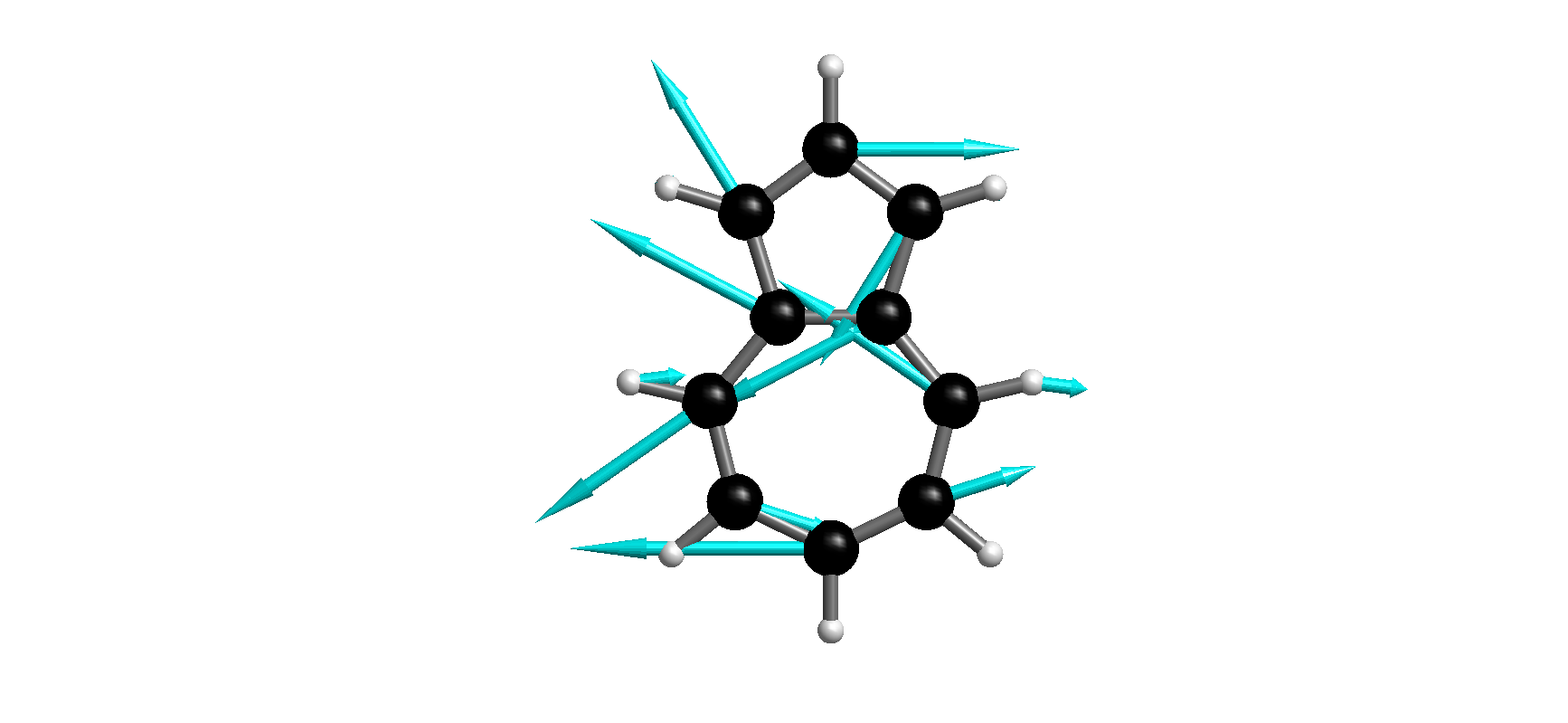}} &
		\resizebox{0.15\textwidth}{!}{\includegraphics[angle=90, trim=600 0 600 0, clip]{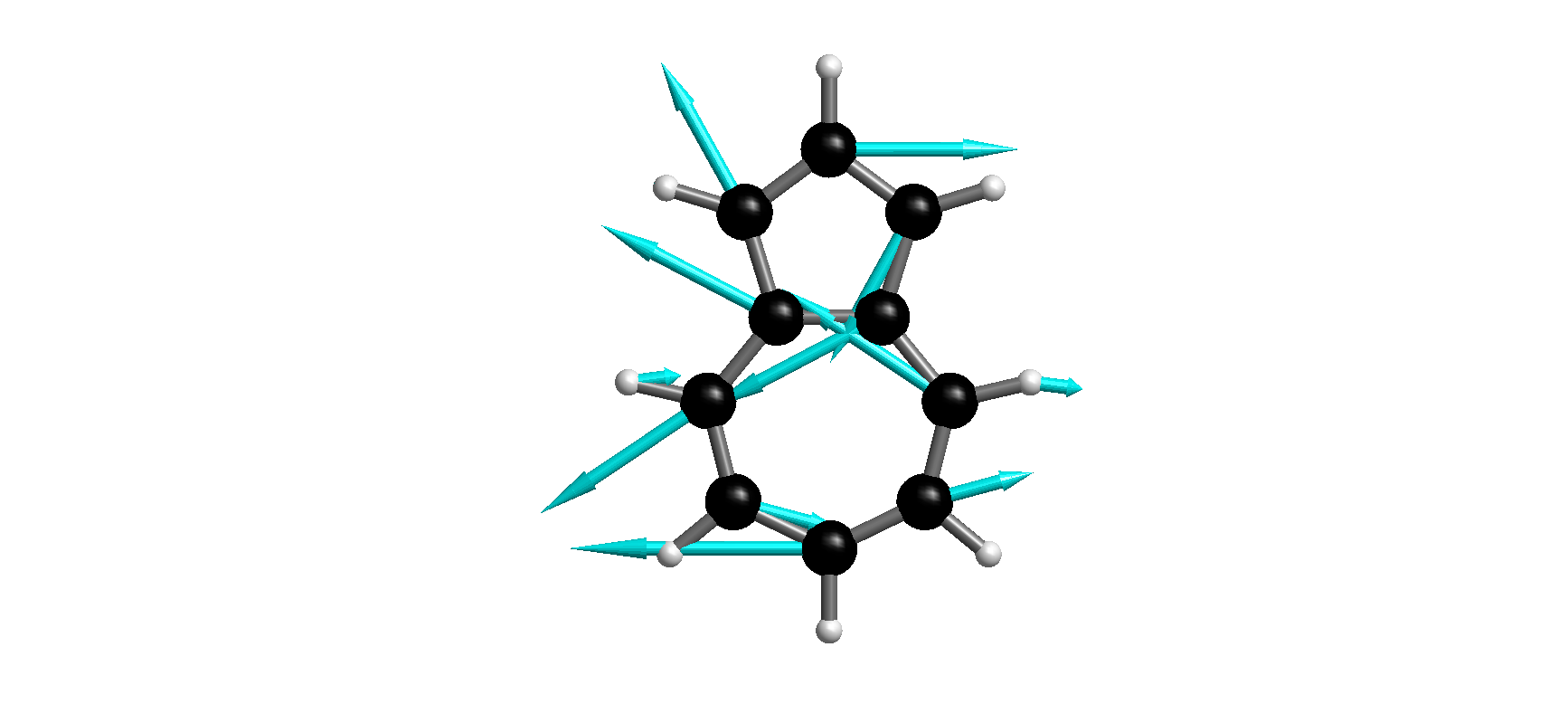}} &
		\resizebox{0.15\textwidth}{!}{\includegraphics[angle=90, trim=600 0 600 0, clip]{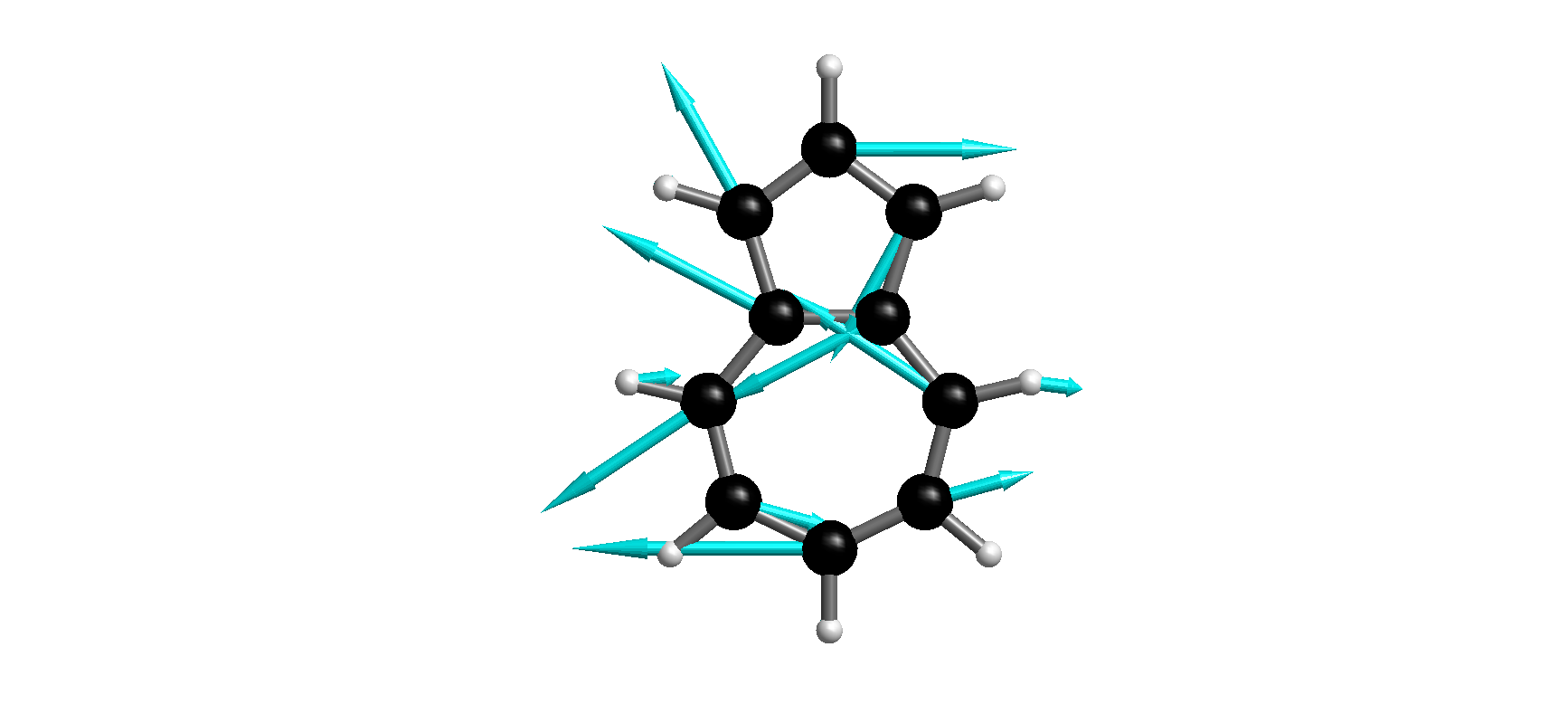}} \\
		
		$g_{T_1T_2}^{\xi, \mathrm{CBS(TDPT)}}$&
		\resizebox{0.15\textwidth}{!}{\includegraphics[angle=90, trim=600 0 600 0, clip]{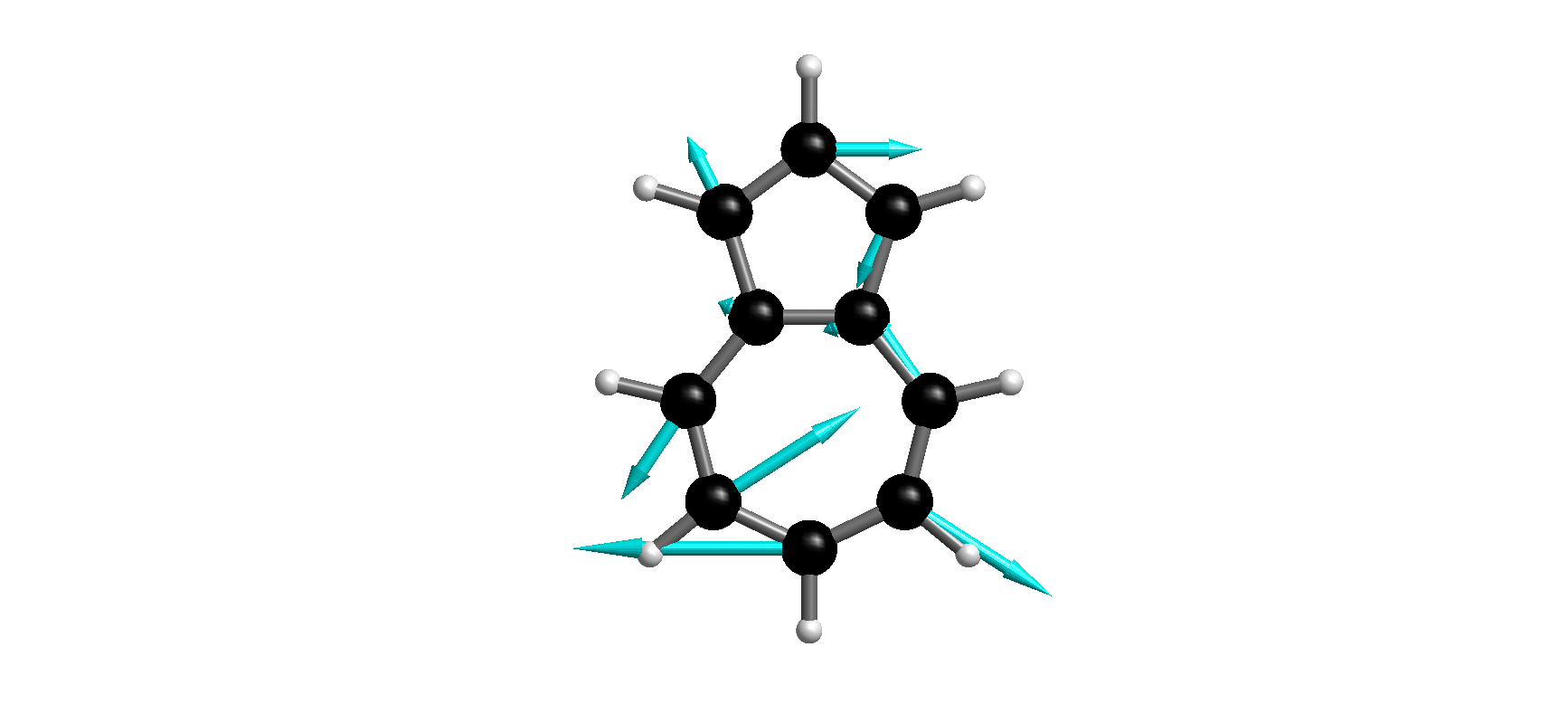}} &
		\resizebox{0.15\textwidth}{!}{\includegraphics[angle=90, trim=600 0 600 0, clip]{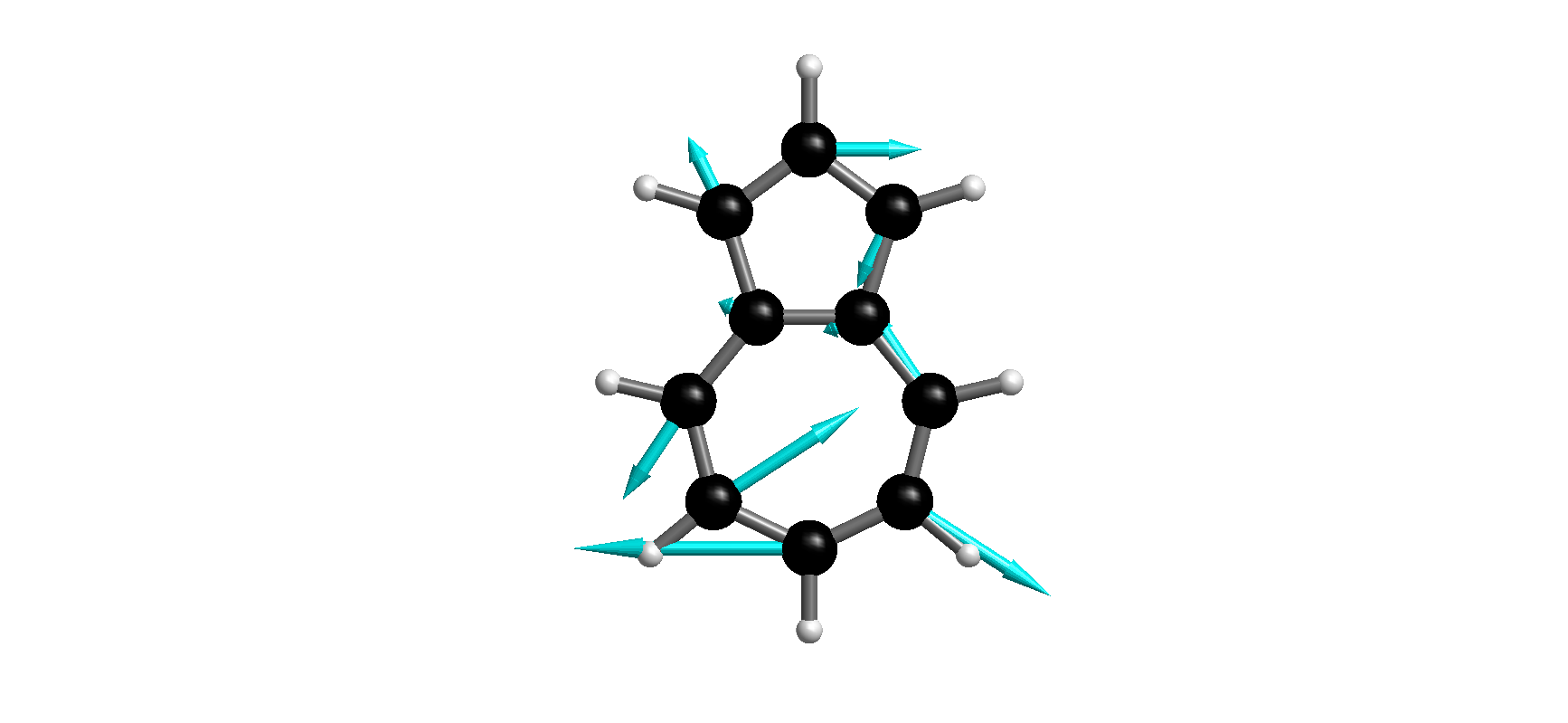}} &
		\resizebox{0.15\textwidth}{!}{\includegraphics[angle=90, trim=600 0 600 0, clip]{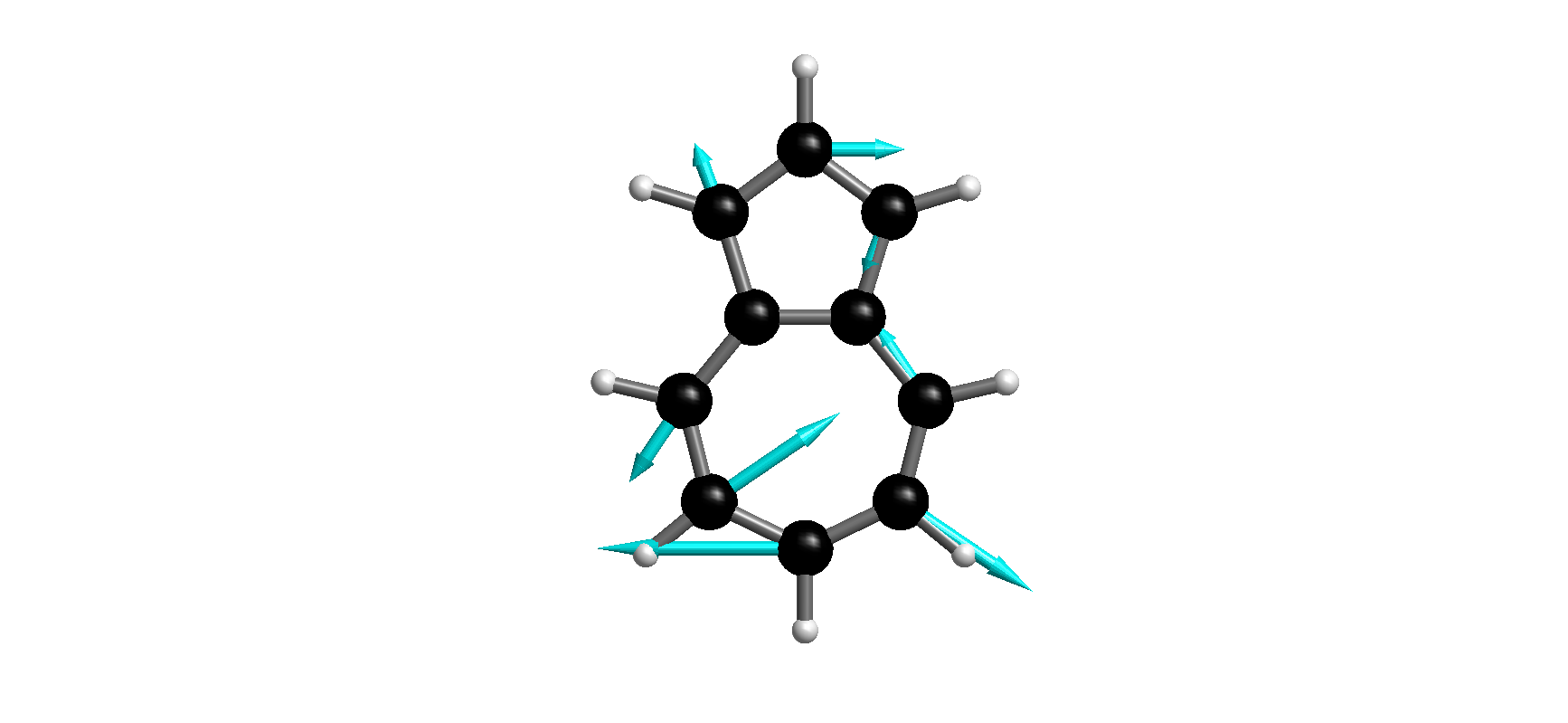}} &
		\resizebox{0.15\textwidth}{!}{\includegraphics[angle=90, trim=600 0 600 0, clip]{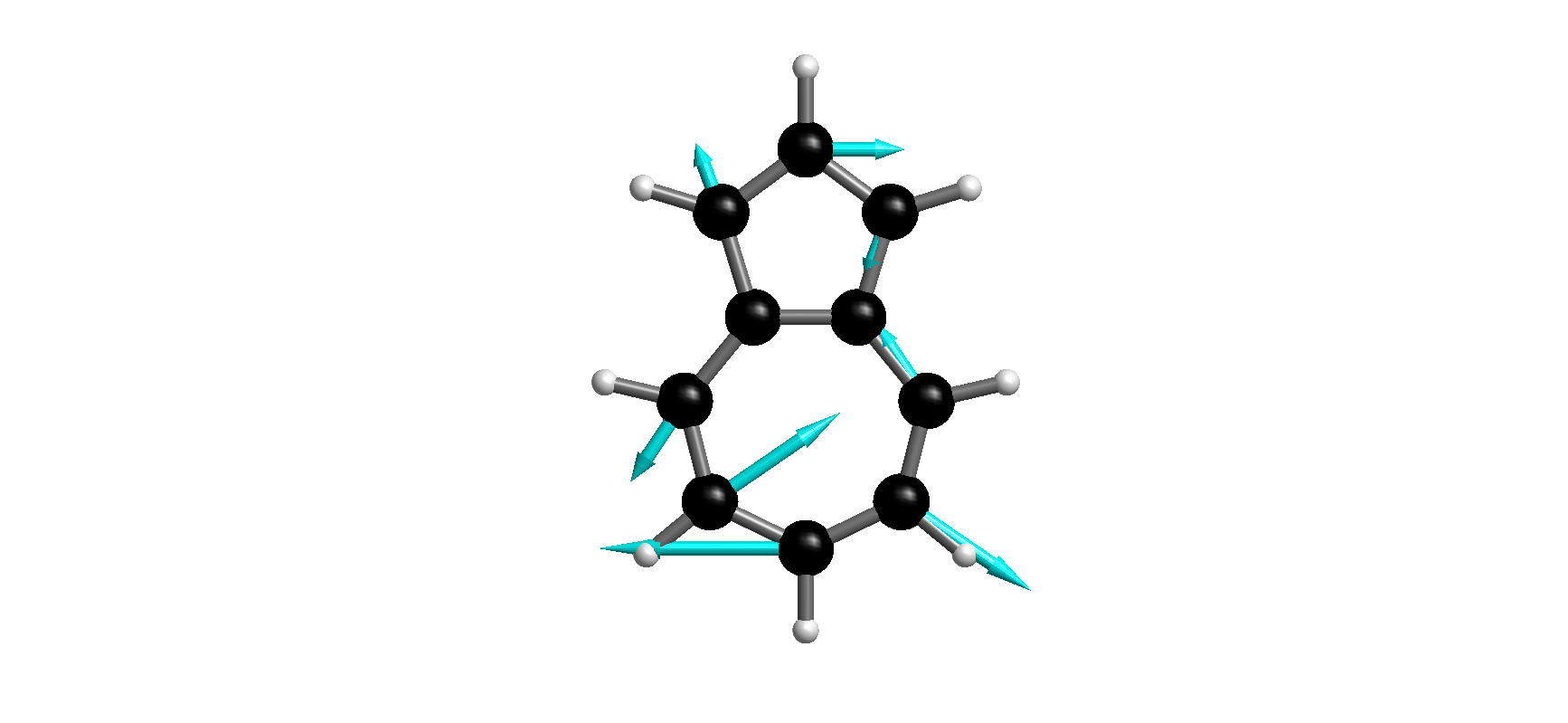}} \\
	\end{tabular}
	\caption{The $S_0$-$S_1$, $S_1$-$S_2$, and $T_1$-$T_2$ fo-NACME vectors of azulene calculated by TDDFT/B3LYP with different basis sets
(NB: the $T_1$-$T_2$ vectors are scaled by a factor of 1/3). For additional explanations, see Table \ref{RMSD}.} \label{fig:nacme}
\end{figure}

The first thing to note is the basis set convergence rates of the three methods. To this end, the def2-SV(P), def2-SVP, def2-TZVP, and def2-TZVPP  basis sets\cite{def2} were used,
with the results documented in Table \ref{RMSD}. As can be seen, the root-mean-square (RMS) deviations
of the EOM and TDPT fo-NACMEs from the near CBS results are smaller by  2 to 3 orders of magnitude than those by Eq. \eqref{HFey_rho}.
Notably, while $g^{\xi,\mathrm{CBS(TDPT)}}$ approaches slowly $g^{\xi,\mathrm{TDPT}}(\textrm{def2-TZVPP})$ as the basis set is enlarged,
$g^{\xi,\mathrm{CBS(EOM)}}$ does not approach $g^{\xi,\mathrm{EOM}}(\textrm{def2-TZVPP})$ and grossly overestimates the fo-NACMEs
compared with both $g^{\xi,\mathrm{EOM}}$ and $g^{\xi,\mathrm{TDPT}}$ (cf. Fig. \ref{fig:nacme}).
In fact, it can be proven\cite{lzdNACtheory} that $g^{\xi,\mathrm{CBS(TDPT)}}$ (for \emph{ee} fo-NACMEs) coincides with $g^{\xi,\mathrm{TDPT}}$, but
$g^{\xi,\mathrm{CBS(EOM)}}$ in general does not agree with $g^{\xi,\mathrm{EOM}}$ in the CBS limit. Therefore, from both numerical and theoretical points of view,
$g^{\xi,\mathrm{CBS(EOM)}}$ should not be recommended as a useful approach.

The overall accuracy of the calculated fo-NACME can be assessed by their use in computing the IC rate constants (at $T=300$ K)
of $S_1 \to S_0$, $S_2 \to S_1$, and $T_2 \to T_1$. The MOMAP software package\cite{momap, peng2007excited, niu2008promoting} was used here,
with Duschinsky rotation effects accounted for. The harmonic vibrational frequencies were derived from the numerical Hessian yet based on
analytically evaluated TDDFT/B3LYP/def2-SVP gradients\cite{XTDDFTgrad} available in BDF.
The results are collected in Table \ref{kIC}, to be compared with
the experimental values ($5.3\times 10^{11}$ s$^{-1}$ for
$S_1 \to S_0$\cite{azuleneS1IC} and $3.5\times 10^8$ s$^{-1}$ for $S_2 \to S_1$\cite{azuleneS2IC,azuleneS2S0IC}). 
It appears that all methods tend to underestimate the IC rate constant of $S_1 \to S_0$, which undergoes mainly through a conical intersection\cite{azuleneS1S0conical}. However,
this cannot be fully ascribed to the quality of the calculated fo-NACMEs, since the harmonic approximation for the vibrations
also tends to underestimate the rate constant of processes through conical intersections (CX). Such argument is supported by
the fact that the rate constant of the non-CX $S_2 \to S_1$ is well reproduced by both the EOM and TDPT based approaches.
In contrast, both $k_{S_2\to S_1}^{\mathrm{CBS(TDPT)}}$ and especially $k_{S_2\to S_1}^{\mathrm{CBS(EOM)}}$ are in large errors,
showing again that the Hellmann-Feynman-like expression \eqref{HFey_rho} is unreliable.
The $T_2 \to T_1$ IC turns out to be orders of magnitude faster than both $S_2 \to S_1$ and $S_1 \to S_0$,
due to a much smaller energy gap (0.29, 1.34, and 1.59 eV for $T_2$-$T_1$, $S_2$-$S_1$, and $S_1$-$S_0$, respectively). Because of this,
 the $T_2 \to T_1$ IC cannot readily be detected experimentally.


\section{Conclusions and outlook}\label{sec:conclusion}
Different formulations of NAC-TDDFT for the fo-NACMEs between the ground and excited states and between two excited states
have been analyzed critically and demonstrated numerically. The take-home messages include
(1) the Hellmann-Feynman type of formulation is hardly useful due to huge demands on basis sets;
(2) the AWF-based formulation is not theoretically founded and may give incorrect expressions; (3) the TDPT-based formulation is most rigorous but
the quadratic responses therein may be problematic, not because of the formulation itself but due to the approximate nature of the exchange-correlation kernel;
(4) the EOM-based formulation is not only elegant but is also free of numerical instabilities, and is therefore the recommended choice.
Further extensions of the EOM variant of NAC-TDDFT include (a) use of fragment localized molecular orbitals\cite{FLMO,Triad,ACR-FLMO,FLMO3} for
efficient calculations of large systems; (b) combination with spin adaptation\cite{SARPA,SATDDFT,XTDDFT,XTDDFTgrad} for proper treatment of open-shell systems; 
(c) combination with perturbative treatment of SOC\cite{X2CSOC1,X2CSOC2,X2CTDSOC}, which is another source for nonadiabatic couplings; and
(d) combination with variational treatment of SOC\cite{RTDDFT1,RTDDFT2,RTDDFT3,X2CTDYbO,X2CTDReO4,X2CTDOsO4,RTDDFTrev,LiuPerspective2020} for systems containing very heavy elements. Progresses along these directions are being made at our laboratory.

\subsection*{\sffamily \large ACKNOWLEDGMENTS}
This work was supported by the National Key R\&D Program of China (Grant No. 2017YFB0203402),
National Natural Science Foundation of China (Grant Nos. 21833001 and 21973054), Mountain Tai Climb Program of Shandong Province,
and Key-Area Research and Development Program of Guangdong Province (Grant No. 2020B0101350001).

\subsection*{\sffamily \large BIOGRAPHICAL INFORMATION}
\textbf{Z. Wang} was born in 1995 and obtained his Ph. D. in Chemistry from Peking University (2018).
His research interests include spin-adapted open-shell time-dependent density functional theory, fragmentation and semiempirical methods.

\noindent
\textbf{C. Wu} was born in 1994 and obtained his Ph. D. in Chemistry from University of New South Wales (2020).
His research focuses on computer-aided rational design of organic dyes and photocatalysts for polymerization.

\noindent
\textbf{W. Liu} is Chair Professor at Shandong University, member of International Academy of Quantum Molecular Science, fellow of Royal Society of Chemistry
and Asia-Pacific Association of Theoretical and Computational Chemists, recipient of
Bessel Award of Humboldt Foundation, Annual Medal of International Academy of Quantum Molecular Science,
Pople and Fukui Medals of Asia-Pacific Association of Theoretical and Computational Chemists.
His research interests include quantum electrodynamics, relativistic electronic structure theory, many-body theory of strong correlation,
time-dependent density functional theory, etc.

\bibliography{BDFlib}

\clearpage
\newpage

\begin{table}
	\caption{Root-mean-square deviations of TDDFT/B3LYP fo-NACMEs (in Bohr$^{-1}$) between different states of azulene from the reference data:
 $g^{\xi, \mathrm{EOM}}(\textrm{def2-TZVPP})$ for $g^{\xi, \mathrm{EOM}}$ and $g^{\xi, \mathrm{CBS(EOM)}}$, and $g^{\xi, \mathrm{TDPT}}(\textrm{def2-TZVPP})$ for $g^{\xi, \mathrm{TDPT}}$ and $g^{\xi, \mathrm{CBS(TDPT)}}$.
  $g^{\xi,\mathrm{EOM}}$ by Eq. \eqref{NAC_EOM_0I}/\eqref{NAC_EOM_IJ}; $g^{\xi,\mathrm{TDPT}}$ by Eq. \eqref{NAC_EOM_0I}/\eqref{RTee}; $g^{\xi, \mathrm{CBS(EOM)}}$ by Eq. \eqref{HFey_rho} with
  $\tilde{\boldsymbol{\gamma}}^{IJ}$ \eqref{ee-gammaEOM}; $g^{\xi, \mathrm{CBS(TDPT)}}$ by Eq. \eqref{HFey_rho} with $\bar{\boldsymbol{\gamma}}^{IJ}$ \eqref{TDPTrTDM}.}
	\begin{tabular}{lcccc}
		\hline\hline
		& def2-SV(P) & def2-SVP & def2-TZVP & def2-TZVPP \\
		\hline
		$g_{S_0S_1}^{\xi, \mathrm{EOM}}$& 0.0037&	0.0006&	0.0002&	0.0000
		\\
		$g_{S_0S_1}^{\xi, \mathrm{CBS(EOM)}}$& 0.1077&	0.1072&	0.0601&	0.0585
		\\
		\hline
		$g_{S_1S_2}^{\xi, \mathrm{EOM}}$& 0.0028&	0.0031&	0.0003&	0.0000
		\\
		$g_{S_1S_2}^{\xi, \mathrm{TDPT}}$& 0.0037&	0.0042&	0.0003&	0.0000
		\\
		$g_{S_1S_2}^{\xi, \mathrm{CBS(EOM)}}$& 0.4774&	0.4746&	0.4949&	0.4941
		\\
		$g_{S_1S_2}^{\xi, \mathrm{CBS(TDPT)}}$& 0.0865&	0.0841&	0.0523&	0.0505
		\\
		\hline
		$g_{T_1T_2}^{\xi, \mathrm{EOM}}$& 0.0726&	0.0775&	0.0039&	0.0000
		\\
		$g_{T_1T_2}^{\xi, \mathrm{TDPT}}$& 0.0724&	0.0774&	0.0039&	0.0000
		\\
		$g_{T_1T_2}^{\xi, \mathrm{CBS(EOM)}}$& 1.5479&	1.5688&	1.4715&	1.4685
		\\
		$g_{T_1T_2}^{\xi, \mathrm{CBS(TDPT)}}$& 0.4406&	0.4356&	0.2514&	0.2419
		\\
		\hline\hline
	\end{tabular}
	\label{RMSD}
\end{table}

\clearpage
\newpage


\begin{table}
	\caption{TDDFT/B3LYP internal conversion rate constants ($k$ in s$^{-1}$) between different states of azulene.
The experimental values are $5.3\times 10^{11}$ s$^{-1}$ for
$k_{S_1\to S_0}$ \cite{azuleneS1IC} and $3.5\times 10^8$ s$^{-1}$ for $k_{S_2\to S_1}$\cite{azuleneS2IC,azuleneS2S0IC}.
For additional explanations, see Table \ref{RMSD}. }
	\begin{tabular}{lcccc}
		\hline\hline
		& def2-SV(P) & def2-SVP & def2-TZVP & def2-TZVPP \\
		\hline
		$k_{S_1\to S_0}^{\mathrm{EOM}}$& 7.96E+08&	8.00E+08&	7.67E+08&	7.68E+08
		\\
		$k_{S_1\to S_0}^{\mathrm{CBS(EOM)}}$& 1.27E+09&	1.32E+09&	1.05E+09&	1.04E+09
		\\
		\hline
		$k_{S_2\to S_1}^{\mathrm{EOM}}$& 1.29E+09&	1.27E+09&	1.25E+09&	1.25E+09
		\\
		$k_{S_2\to S_1}^{\mathrm{TDPT}}$& 1.38E+09&	1.36E+09&	1.35E+09&	1.34E+09
		\\
		$k_{S_2\to S_1}^{\mathrm{CBS(EOM)}}$& 1.25E+10&	1.26E+10&	1.31E+10&	1.31E+10
		\\
		$k_{S_2\to S_1}^{\mathrm{CBS(TDPT)}}$& 2.25E+09&	2.44E+09&	1.96E+09&	1.94E+09
		\\
		\hline
		$k_{T_2\to T_1}^{\mathrm{EOM}}$& 1.40E+13&	1.42E+13&	1.23E+13&	1.22E+13
		\\
		$k_{T_2\to T_1}^{\mathrm{TDPT}}$& 1.39E+13&	1.41E+13&	1.22E+13&	1.21E+13
		\\
		$k_{T_2\to T_1}^{\mathrm{CBS(EOM)}}$& 8.75E+13&	8.92E+13&	8.14E+13&	8.11E+13
		\\
		$k_{T_2\to T_1}^{\mathrm{CBS(TDPT)}}$& 2.33E+13&	2.33E+13&	1.78E+13&	1.74E+13
		\\
		\hline\hline
	\end{tabular}
	\label{kIC}
\end{table}

\clearpage
\newpage


\end{document}